\documentclass[sigconf]{acmart}

\copyrightyear{2025}
\acmYear{2025}
\setcopyright{acmlicensed}
\acmConference[CHI '25]{CHI Conference on Human Factors in Computing Systems}{April 26-May 1, 2025}{Yokohama, Japan} \acmBooktitle{CHI Conference on Human Factors in Computing Systems (CHI '25), April 26-May 1, 2025, Yokohama, Japan} \acmDOI{10.1145/3706598.3714179}
\acmISBN{979-8-4007-1394-1/25/04}

\usepackage{graphicx}
\usepackage{booktabs}
\usepackage{subcaption}
\usepackage{amsmath}
\usepackage{textcomp}
\newcommand{\texttildenew}{\raisebox{0.5ex}{\texttildelow}}

\begin{document}

\title{HaloTouch: Using IR Multi-path Interference to Support Touch Interactions With General Surfaces}

\author{Ziyi Xia}
\affiliation{%
  \institution{University of British Columbia}
  \city{Vancouver, BC}
  \country{Canada}}
\email{zxia0101@cs.ubc.ca}

\author{Xincheng Huang}
\affiliation{%
  \institution{University of British Columbia}
  \city{Vancouver, BC}
  \country{Canada}}
\email{xchuang@cs.ubc.ca}

\author{Sidney S Fels}
\affiliation{%
  \institution{University of British Columbia}
  \city{Vancouver, BC}
  \country{Canada}}
\email{ssfels@ece.ubc.ca}

\author{Robert Xiao}
\affiliation{%
  \institution{University of British Columbia}
  \city{Vancouver, BC}
  \country{Canada}}
\email{brx@cs.ubc.ca}

\renewcommand{\shortauthors}{Xia et al.}

\begin{abstract}
  Sensing touch on arbitrary surfaces has long been a goal of ubiquitous computing, but often requires instrumenting the surface. Depth camera-based systems have emerged as a promising solution for minimizing instrumentation, but at the cost of high touch-down detection error rates, high touch latency, and high minimum hover distance, limiting them to basic tasks. We developed HaloTouch, a vision-based system which exploits a multipath interference effect from an off-the-shelf time-of-flight depth camera to enable fast, accurate touch interactions on general surfaces. HaloTouch achieves a 99.2\% touch-down detection accuracy across various materials, with a motion-to-photon latency of 150 ms. With a brief (20s) user-specific calibration, HaloTouch supports millimeter-accurate hover sensing as well as continuous pressure sensing. We conducted a user study with 12 participants, including a typing task demonstrating text input at 26.3 AWPM. HaloTouch shows promise for more robust, dynamic touch interactions without instrumenting surfaces or adding hardware to users.
\end{abstract}

\ccsdesc[500]{Human-centered computing~Interaction techniques}

\keywords{Input Techniques, Touch Sensing, Ubiquitous Computing, Pressure and Proximity Sensing, Ad-hoc Surface Interaction, On-world Interaction}

\maketitle

\section{Introduction}
As computing becomes increasingly ubiquitous, it is moving beyond the confines of small screens and dedicated devices. This evolution is paving the way for ``on-world'' interfaces that leverage the physical environment itself for interaction \cite{Xiao2018OnWorld, Huang2024SurfShare}. Traditional input methods, such as mice and keyboards, though precise, are tied to specific devices and lack the flexibility required for spontaneous, ad-hoc use in dynamic settings. Recent advances in natural input techniques, like hand gestures  \cite{Vatavu2023Gestures, Hinckley2016PreTouchSF, Csonka2023AIGesture} and speech recognition \cite{Kamel1990, Bolt1980VoiceGesture, Schmandt1982VoiceInterface}, have enabled more intuitive and controller-free interactions. However, these methods are often limited by a lack of haptic feedback \cite{Lindeman1999TowardsVR, Meier2021TapID}, reduced precision \cite{Cheng2022Comfortable, Medeiros2023Beniefits}, and user fatigue~\cite{Brasier2020ARPads, Hincapi2014MidAir} over extended periods. To overcome these challenges, the focus is now on developing robust input sensing techniques that can transform everyday surfaces into interactive, touch-sensitive interfaces. 

\begin{figure*}[htbp!]
  \centering  \includegraphics[width=1.0\linewidth]{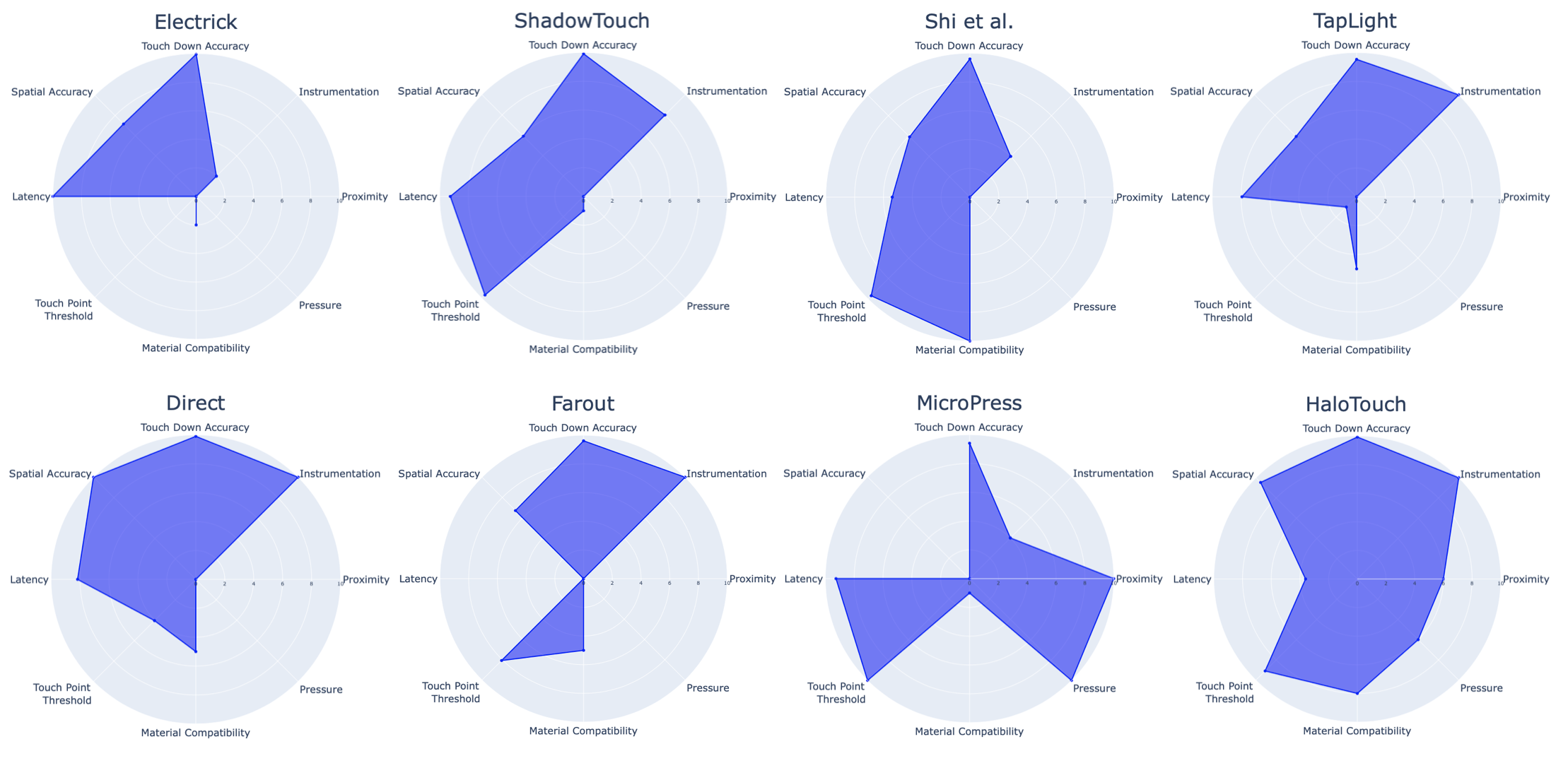}
  \caption{Radar charts comparing eight touch input systems (Electrick\cite{Zhang2017Electrick}, ShadowTouch\cite{Liang2023ShadowTouch}, Shi et al.\cite{Shi2020FingerIMU}, TapLight\cite{Streli2023LightSpeckle}, Direct\cite{Xiao2016Direct}, FarOut\cite{Shen2021Farout}, MicroPress\cite{Dobinson2022MicroPress}, HaloTouch) across metrics like Accuracy, Latency, Material Compatibility}
  \Description{A set of eight radar charts comparing the performance of Electrick, ShadowTouch, Shi et al., TapLight, Direct, FarOut, MicroPress, and HaloTouch across various evaluation criteria. Each chart has axes representing Touch Down Accuracy, Spatial Accuracy, Latency, Touch Point Threshold, Material Compatibility, Instrumentation, Pressure, and Proximity. The filled blue areas indicate each system’s relative performance. Notably, HaloTouch has strong overall accuracy, spatial precision, and supports pressure and proximity sensing, whereas other systems excel in specific areas like latency or material compatibility.}
  \label{fig:radar_graph}
\end{figure*}

Several techniques have been proposed to enable touch input on various surfaces. For instance, Electrick \cite{Zhang2017Electrick} uses electric field tomography to achieve remarkably high touch accuracy. Taplight~\cite{Streli2023LightSpeckle} reduces touch latency to an unprecedented 50.4ms by analyzing structured light patterns. ShadowTouch \cite{Liang2023ShadowTouch} has a \emph{touch point threshold} of less than 2mm -- accurately triggering touch down events only when the finger is within 2mm of the surface. Tripad~\cite{Dupre2024Tripad} takes a human-centric approach to generalize touch input across a wide range of surface materials. Meanwhile, Micropress~\cite{Dobinson2022MicroPress} combines pressure and proximity sensing by integrating an IMU (Inertial Measurement Unit) into the fingers. While each of these innovative techniques excels in one or two specific metrics, none of them can comprehensively address the entire design space of touch input (see Figure \ref{fig:radar_graph}).

We present \textbf{HaloTouch}, a novel technique enabling comprehensive sensing input capabilities with a commercially available depth camera. \textbf{HaloTouch} achieves a touch-down accuracy of 99.2\% across 5 different materials, spatial accuracy within 5.5mm, a touch point threshold of 4.97mm, and a motion-to-photon latency of 150ms. Furthermore, the pressure and proximity detection capabilities showed high reliability, with mean pressure error of 18.77\%  and mean proximity error of 2.81 mm.

\textbf{HaloTouch} leverages the underexplored phenomenon of depth camera multipath interference, enabling it to work on general day-to-day surfaces. Multipath interference provides a continuous signal around the fingertip area as the finger approaches the surface, enabling us to determine the fingertip's proximity to the surface despite sensor noise. We refer to this phenomenon as the Halo effect, as it visually appears as a ``halo'' around the fingertip. This technique requires no surface instrumentation or wearable sensors, though we do need to perform a short (20s) one-time calibration for users on a new material. The unique Halo phenomenon also enables touch input in challenging conditions, such as very wet surfaces.

After comparing with other related work, we came up with design requirements for \textbf{HaloTouch}. We describe the theory behind multipath interference, the Halo effect and our approach to model this signal for building a touch detection pipeline. We then report on a two part study, the first one capturing the technical evaluations of \textbf{HaloTouch} and the second one showing the interaction modalities it enables with passive surfaces.

Finally, we summarize our contributions as follows:

\begin{itemize}
    \item Explored and characterized the Halo effect as a way to surpass noise thresholds in commodity depth sensors.
    \item Developed a comprehensive multimodal touch input system, supporting both pressure and proximity sensing, for use on passive physical surfaces using only a depth+RGB camera.
    \item Demonstrated the versatility and effectiveness of our system for rapid, ad hoc use cases through a user study focused on typing, alongside two additional example applications.
\end{itemize}

\section{Related Work}

Prior approaches, including surface instrumentation, wearables, and vision-based methods, offer trade-offs in accuracy, scalability, and practicality. While depth-sensing and computer vision provide promising alternatives, existing methods often struggle with real-time performance or deployment complexity. This section reviews prior work in ad-hoc touch, pressure, and proximity sensing.

\subsection{Enabling Ad-hoc Touch Input on Passive Physical Surfaces}

Researchers have explored various methods for capturing touch events on everyday surfaces. One approach is surface instrumentation with acoustic sensors or electrodes. For example, prior research has localized touch events on surfaces with signals from contact microphones~\cite{Ishii1999PingPongPlus,Paradiso2002AccousticKnocks,Xiao2014Toffee,Harrison2008Scratch}. Similarly, electrodes can also be added to objects or surfaces to make them touch-sensitive \cite{Zhang2017Electrick, Buechley2010LivingWall, Zhang2018Wall++}. However, instrumenting surfaces introduces deployment overhead and is costly when scaling to additional surfaces. Other research has explored instrumenting users with wearables to detect touch signals. SkinTrack \cite{Zhang2016SkinTrack} integrates four electrodes into a smartwatch, localizing finger touches by measuring phase differences of active electrical signals emitted from a separate finger-worn device. ShadowTouch \cite{Liang2023ShadowTouch} adds a light emitter to a user's wrist and predicts subtle touch events by analyzing the shadow of the user's hand. Additionally, prior research has developed small finger-mounted IMUs to detect touch actions during interactions. Gu et al.~\cite{Gu2019FingerIMU} used an IMU in a finger ring to achieve low latency touch detection, whereas Oh et al.~\cite{Oh2020FingerTouch} mounted the IMU directly on the fingertip for high accuracy touch contact detection. ActualTouch \cite{Shi2020FingerIMU} mounted a small IMU on the user's fingertip to detect differences in finger microvibrations between touching/dragging and in-air states.

Camera-based approaches for sensing ad-hoc touch have become popular as they do not require instrumenting the surface nor the user. Wilson \cite{Wilson2010UsingAD} was the first to demonstrate the possibility of using a depth camera to detect touches on flat surfaces. Other explorations have focused on improving different metrics of touch input, such as providing high touch down accuracy \cite{Xiao2016Direct, Harrison2011Omnitouch}, low spatial error \cite{Cadena2016Fingertip, Haubner2013Singledepth, Kurz2014ThermalTouch}, low latency \cite{Fan2022ReducingLatency, Streli2023LightSpeckle}, and longer sensing range \cite{Shen2021Farout}. Among those, DIRECT \cite{Xiao2016Direct} combined depth and IR signal sources to reach more than 99\% touch down accuracy at the cost of higher latency. TapLight \cite{Streli2023LightSpeckle} leveraged structured laser light reflection disparity to reach a very low latency (50.4~ms) but their touch point threshold is as high as 45~mm.

Deploying ad-hoc touch input systems on headsets provides precise, tactile input for augmented and virtual reality applications~\cite{Huang2024VirtualNexus, Huang2023VR360Telepresence}, complementing existing natural input techniques. MRTouch \cite{Xiao2018MRTouch} achieved real-time surface plane detection and touch input detection using sensors embedded in HoloLens 2. It demonstrated the potential for fast touch input interactions using IR and depth-sensing techniques, although no formal evaluation was conducted to assess how rapid users can give touch inputs to the system. Tripad \cite{Dupre2024Tripad} used a more human-centric approach to generalize touch input to arbitrary surfaces using only hand tracking from HoloLens 2. It successfully performed basic input tasks like clicking and dragging, though the touch point threshold remains relatively high at 55~mm, making it difficult to rapidly disengage touches for fast input. EgoTouch \cite{Mollyn2024EgoTouch} uses Apple Vision Pro's RGB cameras for calibration-free, on-skin touch detection, achieving high accuracy in various environments, but it struggled with high-speed input and occlusions. TouchInsight \cite{Streli2024TouchInsight} employed a probabilistic framework to handle egocentric hand tracking uncertainties, enabling precise, low-latency interactions like virtual keyboard typing. In contrast, \textbf{HaloTouch} combines pressure and proximity sensing with accurate touch input across general surfaces, expanding interaction capabilities.

\subsection{Ad hoc Pressure Sensing and Proximity Sensing}

Traditional pressure sensing methods involve physical sensors such as capacitive grids \cite{Ahuja2021TouchPose, Guo2015CapAuth, Choi2021Touchscreens}, force-sensitive resistors \cite{Brahmbhatt2020Grasp, Pham2018HandObjectCF}, or flexible sensors \cite{bhirangi2021reskin, Kim2011CapacitiveTS}, which must be mounted on surfaces or hands. These methods can be impractical and expensive for diverse, real-world applications. Vision-based surface pressure sensing is a growing field that aims to estimate contact pressure using computer vision techniques rather than invasive physical sensors.

Recent advances in computer vision have allowed for pressure estimation using visual cues such as fingertip color changes \cite{Chen2019EstimatingFF, Mascaro2001PhotoplethysmographFS, Mascaro2004MeasurementOF} and soft tissue deformation \cite{Hwang2017InferringIF}. ForceSight \cite{Pei2022ForceSight} showed high pressure sensing potentials using laser speckle pattern differences reflected from deformed objects, though it only works with plastically deformable materials. PressureVision \cite{Grady2022PressureVision} pioneered using deep learning to estimate fingertip pressure from RGB images, but it was limited by controlled settings with constrained lighting and simple surfaces. To address these limitations, the ``PressureVision++''~\cite{Grady2024PressureVision++} model utilized weak supervision through contact labels. This method requires the collection of diverse training data without requiring ground-truth pressure measurements from high-resolution sensors. In comparison, \textbf{HaloTouch} leverages learning of the IR and depth multipath interference signal, avoiding the use of large training set to estimate a more fine-grained pressure level on flat surfaces.

Pressure sensing and proximity sensing can be achieved at the same time to enable new modes of interaction. MicroPress \cite{Dobinson2022MicroPress} used Inertial Measurement Unit (IMU) sensors combined with deep learning models to detect both pressure and hover distance between fingers with high precision. However, this method requires additional sensors to be placed on both fingers, potentially limiting its practicality for casual or on-the-go use. 
Other existing approaches focus on finger proximity detection in mobile interactions. Air+Touch~\cite{Chen2014AirTouch} employed a horizontally mounted depth camera on a smartphone to explore the interaction space for both touch and hover gestures. Similarly, OmniSense \cite{Yeo2023OmniSense} used a 360-degree camera feed on a mobile phone to train a neural network that predicts the 3D location of fingertips. Another technique proposed by Matulic~\cite{Matulic2023AboveScreen} involved two mirrors mounted above the phone screen that reflect the front camera view; captured video is sent to a deep neural network that robustly infers the 3D position of fingertips. Although effective, these methods still require some level of instrumentation, either on the touch surfaces or the mobile devices, which makes achieving spontaneous, ad hoc proximity sensing difficult. 
\textbf{HaloTouch} uses only a commodity available depth camera, offering the potential to achieve ad hoc proximity sensing and pressure sensing for any surfaces anywhere.

\section{Positioning and Goals of HaloTouch}

We identified representative touch sensing technologies with high reported accuracy (above 95\%) and distilled eight key technical dimensions. From these, we selected works that demonstrated the best performance in each dimension to ensure a comparison against state-of-the-art benchmarks. Finally, we evaluated and mapped the overall performance of the selected works across these eight dimensions. The dimensions and their scoring criteria are detailed below, with additional scoring information provided in Appendix Figure \ref{fig:radar_table}.

\begin{itemize}
    \item \textit{Instrumentation}: Past touch sensing systems are mainly based on 3 types of instrumentation: 1) instrumenting surfaces, 2) wearables, and 3) vision-based (i.e., use headsets or Kinect). To score them, we consider deployment friction, scalability, and interactive freedom. Systems that require instrumenting the surface have a lower score because they require the most effort to deploy and are hard to scale. We assign technologies that instrument users' hands with wearables a medium score as they can be bulky and thus affect interactive freedom. Consequently, we consider vision-based technology to have lower deployment friction and better interactive freedom.
    \item \textit{Touch Down Accuracy}: We extract the touch down accuracy reported from the corresponding literature in percentages and normalize them to 0--10 (0\% has a score of 0 and 100\% has a score of 10).
    \item \textit{Spatial Accuracy}: We extract the spatial accuracy from the corresponding literature in mean errors of millimeters, and min--max normalize them to 0--10. Some work (e.g., ShadowTouch~\cite{Liang2023ShadowTouch}, Gu et al.~\cite{Gu2019FingerIMU}, Shi et al.~\cite{Shi2020FingerIMU}, and TapLight~\cite{Streli2023LightSpeckle}) only focus on touch down accuracy. Such work can be combined with built-in hand tracking for touch localization; we thus assign a typical hand tracking spatial accuracy of 11~mm \cite{Abdlkarim2023MetaQuest} to them for scoring.
    \item \textit{Latency}: We extract the latency from the corresponding literature in milliseconds, and min--max normalize them to 0--10.
    \item \textit{Touch Point Threshold}: We define the touch point threshold as the minimum distance the fingertip must travel perpendicular to the surface to register a valid input, while avoiding false positives. We extract the touch point threshold from the corresponding literature in millimeters, and min--max normalize them to 0--10. For work that did not report a specific touch point threshold (i.e., Electrick~\cite{Zhang2017Electrick}, Gu et al.~\cite{Gu2019FingerIMU}, Shi et al.~\cite{Shi2020FingerIMU}), we estimated it from their technical principle and demo video.
    \item \textit{Material Compatibility}: This is an under-explored yet important dimension. Our score is based on our interpretation of their technical principle. For example, technologies that require conductive material~\cite{Zhang2017Electrick} or plain texture of surfaces~\cite{Liang2023ShadowTouch} have a lower material compatibility. 
    \item \textit{Pressure}: While novel track pads \cite{Rendl2014Presstures} already support pressure sensing for touch, enabling the same on everyday surfaces is scarce in literature. 
    \item \textit{Proximity}: For our evaluation, we selected a recent high-performance system \cite{Dobinson2022MicroPress} as a benchmark and assessed our system based on their proximity accuracy metrics.

\end{itemize}

We compare past touch-sensing research with their performance in common metrics distilled from literature. We select representative research that falls in the instrumentation categories we have reviewed: 1) instruments the surface with sensors, 2) instruments the user with wearables, and 3) instrumentation free (besides using RGB/Depth sensing equipment). State-of-the-art touch sensing has reached 99\% accuracy and millimeter-level spatial accuracy while having a low latency. However, few (only ShadowTouch~\cite{Liang2023ShadowTouch} to the best of our knowledge) have focused on achieving a small touch point threshold, enabling subtle touch detection. With \textbf{HaloTouch}, we set out to minimize the touch point threshold while achieving similar touch-event and spatial accuracy to the state-of-the-art, enabling ad hoc, rapid touch interactions on passive physical surfaces. As such, we list the technical requirements for our system:

\begin{enumerate}
    \item Avoid instrumenting users or the surface with additional hardware, thus supporting better interaction freedom and low-friction deployment.
    \item Achieve high accuracy in detecting touch-down moments to ensure that every intentional touch by the user is correctly identified and registered by the system.
    \item Provide high spatial accuracy in touch detection, allowing the system to precisely identify the exact location of each touch input.
    \item Ensure the interaction system responds to touch inputs with a low latency to maintain a fluid and responsive user experience.
    \item Minimize the touch point threshold to allow for closely spaced touch inputs without interference. This feature is particularly important in scenarios where rapid or complex gestures are performed, such as fast tapping on multiple locations on the surface.
    \item Support operation across diverse surface materials to enhance usability in various mobile environments.
\end{enumerate}

\textbf{HaloTouch} builds upon the under-explored phenomenon of multipath interference in depth cameras, first highlighted by FarOut \cite{Shen2021Farout}. While FarOut utilized this phenomenon along with additional engineering techniques to extend touch detection range up to 3 meters, our work provides a more detailed explanation of the underlying principles and a comprehensive characterization of multipath interference for touch sensing with depth cameras. Furthermore, our system extends the capabilities of touch input beyond mere contact detection to include hover and pressure inputs, leveraging multipath interference to enable these additional interaction dimensions.

\section{\textbf{HaloTouch}}
In this section, we first explain the relationship between multipath interference and the Halo effect, and the verification for our theory. Then we detail how we implement our systems to take advantage of the Halo effect to achieve touch sensing.
\begin{figure}[htbp!]
  \centering
  \includegraphics[width=1.0\linewidth]{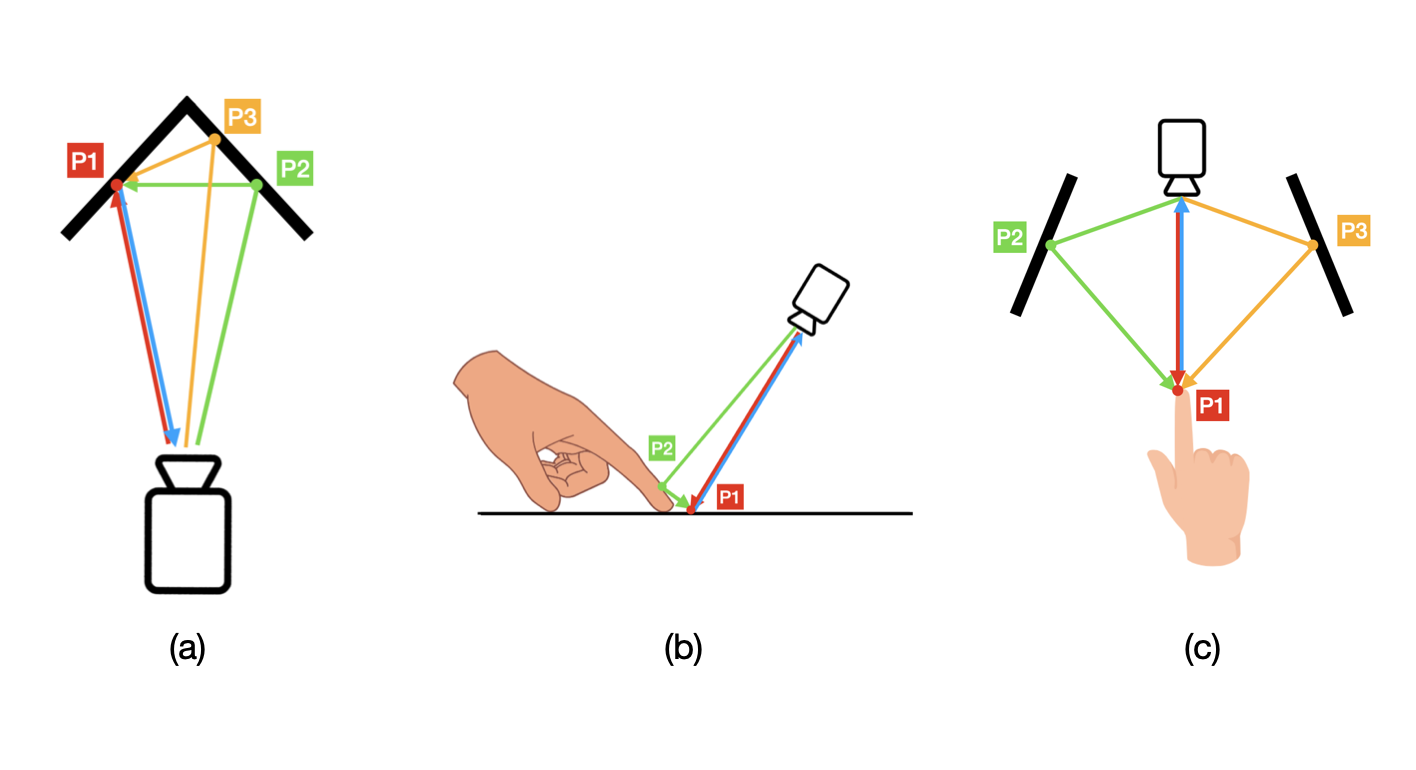}
  \caption{Multi-path Interference example scene: (a) Top view of Multi-path interference when camera faces a corner. (b) Side view of multi-path interference when camera faces a hand. (c) Top view of multi-path interference when camera faces a hand.}
  \Description{A diagram with three labeled subfigures (a, b, c) illustrating multipath interference in depth sensing. (a) shows a camera detecting multiple reflected light paths (P1, P2, P3) from a slanted surface. (b) depicts a fingertip touching a flat surface, with a side-mounted camera receiving direct and indirect reflections. (c) presents a front-view setup where a camera detects a fingertip with reflections bouncing off side surfaces, highlighting the complexity of multipath interference. Colored arrows represent different signal paths contributing to depth sensing variations.}
  \label{fig:multipath_scene}
\end{figure} 

\subsection{Working Principal}

\subsubsection{\textbf{Multipath Interference in Time-of-flight Depth Camera}}
Modern time-of-flight depth cameras rely on correlation techniques to determine the phase shift, and thereby the distance between two signals: a modulated infrared signal that illuminates the scene and the signal received by the sensor, which captures the reflection from the scene's geometry \cite{Lange2001TOFCamera}. Unlike other depth sensing techniques, the time-of-flight depth camera contains an active IR light source that emits a cone of IR light and receives the reflected IR signal using an sensor array to produce a depth map of a 3D scene.

Due to the diffuse reflection pattern of most objects in our life, the same incident light ends up creating multiple light paths while hitting objects with complex geometries. This causes several signals with different path lengths to converge on the same pixel, interfering with each other. Since the sum of two phase-shifted sine waves is itself a phase-shifted sine wave, this interference leads the camera to register incorrect depth measurements, potentially resulting in significant errors in both depth and amplitude – the multipath interference \cite{Feigin2016ReolvingMPI}. 

We explain the multipath interference phenomena in Figure \ref{fig:multipath_scene}(a), when a time-of-flight depth camera illuminates an IR light (red line) into a corner. After the IR signal hits the first contact point (P1), it reflects in many directions following a combination of specular reflection (not drawn) and diffuse reflection (blue line) \cite{Cook1982RmodelCG}. At the same time, the other incident lights hit point P3, creating specular reflection (green line) and hit point P2, creating diffuse reflection (yellow line). Their reflection paths all converge back to the same sensor array cell corresponding to P1, resulting in different phase shifted signals, and thus different computed distances after each signal is correlated with the reference signal. We note that there are typically multiple diffuse reflection sources that converge to P1 in practice, and they all contribute to the final signal strength observed at P1. Therefore, the multipath interference phenomenon generates a complex mixed signal influenced by multiple factors. Additionally, for smooth surfaces that only create specular reflections, the multipath interference phenomena will not be observed since each contact point (P1) will only receive one unique specular reflection coming from the reflection point (P3).

\subsubsection{\textbf{Multipath Interference in Touch Down Events}}
Next, we explain the multipath interference phenomenon in the context of a finger touch-down event, influenced by both local and global impacts. First, when a finger approaches the surface closely (within 3~cm), the finger and the target surface form a mini corner, creating local interference (Figure  \ref{fig:multipath_scene}(b)). At this corner, multipath interference occurs in the same manner as previously described: the red line represents the incident light, which results in diffuse reflections along the blue line after hitting point P1. The other incident light ray (green line) hits point P2 and converges with the blue line path to form multipath interference (reflection angle exaggerated in the figure).

In addition to this local impact, we illustrate the global impact from a top-down perspective of the same finger touch-down event (Figure \ref{fig:multipath_scene}(c)). Beyond the single incident light ray (red line) directed at point P1, the infrared light source also emits other rays in a cone shape. We consider two rays (green line and yellow line) at the widest angles of this cone, which are directed at points P2 and P3—representing two points on other objects in the scene. In this scenario, diffuse reflections are the most common, as the objects may not have perfectly angled surfaces to produce specular reflections. The reflected rays eventually also reach P1, causing P1 to receive phase-shifted light originating from other objects in the scene. Both local and global influences contribute to the final signal strength observed.

\subsubsection{\textbf{The Halo Effect}}
The direct impact of multipath interference in many cases takes the form of unbiased noise, which is hard to eliminate. This means that even if we have a higher resolution depth camera, we are limited in our effective depth resolution. Previous researchers had used different techniques to decouple each phase shifted signal to recover the true phase shift; however they have been limited by bandwidth and SNR \cite{Feigin2016ReolvingMPI}. Instead of modeling the formation of this phenomena, we leverage a case of biased multipath interference as observed directly in the output signal in the depth map, after the interference is processed by the camera sensor.

The Halo effect is influenced by several factors, including the material of the reflecting surfaces, the angle between the ground surface and the camera, the geometry of the objects, their orientation relative to the camera, and the distance between them, among others. We note that the Halo effect signal is highly sensitive; even slight deformations of the fingertip on the surface can cause noticeable variations in the Halo signal, whereas the changes in the depth data at the fingertip itself are imperceptible due to noise. Our pressure sensing technique leverages this high level of sensitivity. To model the Halo effect, we considered the index finger's six degrees of freedom, incorporating the effects of material properties and finger deformation when pressure is applied to the finger press. We empirically characterized the Halo effect using a single index finger interacting with a surface while isolating each of the parameters mentioned above. Each parameter exhibited a non-linear relationship with the Halo effect signal strength, as expected, since changes in one parameter often lead to multiple concurrent effects. For instance, altering only the finger's x-position while keeping other parameters constant affects both the distance and orientation between the finger and the camera. Given this non-linear relationship, we adopt a machine learning approach to predict the Halo effect signal strength, as described later in Section \ref{sec:Halo_revealer}.

The key observation that supports our multipath interference hypothesis is that, after extracting the output signal from the depth camera, visualizing it on the depth map, and subtracting the background depth, the resulting values are positive. This indicates that the depth camera interprets these points as being further away than the surface actually is; in other words, the camera perceives a ``denting'' of the surface. Furthermore, this value increases as more interference is introduced into the scene, which aligns with our hypothesis that a mixture of longer-path, phase-shifted signals causes the camera to interpret the pixel as being further away than its actual distance. More technical details are provided in Section \ref{sec:Halo_revealer}.

We considered other potential explanations for this phenomenon, such as signal averaging and over-exposure artifacts. Signal averaging happens when objects are very close to a surface, causing the boundary between them to blur or disappear, but we ruled this out because the phenomenon we observed occurs after signal averaging in the sequence of events. As the fingertip approaches the surface, an additional output pixel signal appears around it when it gets within \texttildenew 2 cm, following an initial signal loss at \texttildenew 3 cm due to averaging. Over-exposure artifacts, which distort the depth map on overexposed infrared images, were also considered. However, unlike our observed phenomenon, which only occurs when an object is near the surface, over-exposure artifacts affect a larger area of the scene regardless of object proximity.

\subsection{Hardware}
\textbf{HaloTouch} is implemented using Microsoft Kinect Azure depth camera \cite{azurekinectdk2019}, a Optoma LDMLTUZST LED projector, and a commercial grade laptop with Intel i7 11800H CPU at 2.3 GHz, and NVIDIA GeForce RTX 3050 Ti GPU. The depth camera is USB connected to the PC and the projector is connected to the PC via HDMI. Each hardware component in our setup is replaceable by other similar off-the-shelf products.

Although there are many types of depth camera sensors, our approach is exclusive to the time-of-flight sensing modality. Such sensors are commonplace, e.g. in the Microsoft Kinect 2, Azure Kinect, and Microsoft Hololens series of devices, and the Halo effect has been observed in each of these devices.

\begin{figure*}[htbp!]
  \centering
  \includegraphics[width=1.0\linewidth]{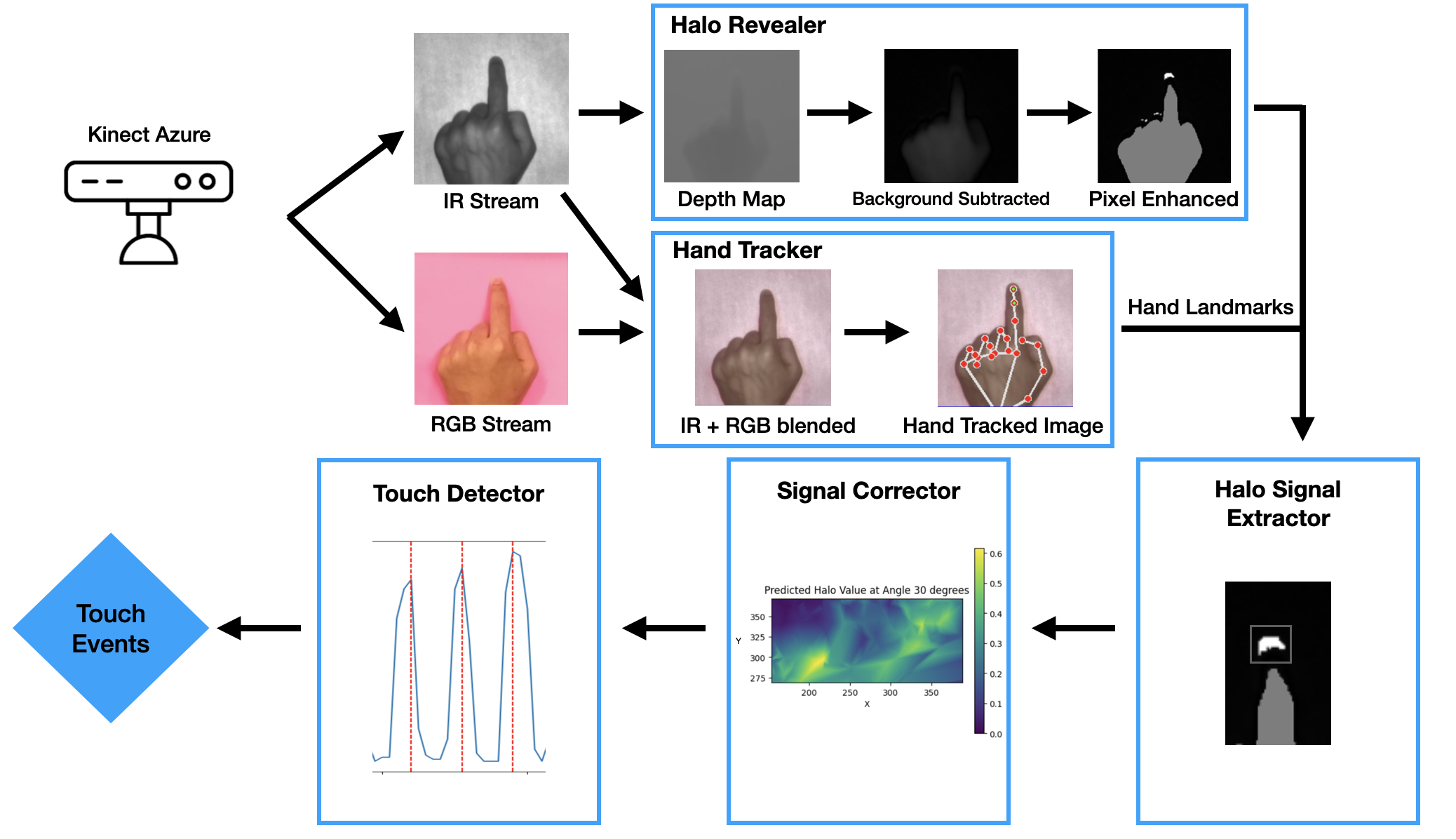}
  \caption{Software: a touch detection pipeline using a Kinect Azure depth camera. It integrates IR and RGB streams with hand tracking and signal processing to isolate fingertip signals, correct noise, and detect touch events}
  \Description{The figure illustrates a pipeline for detecting touch events using a Kinect Azure depth camera, integrating IR and RGB streams, hand tracking, and signal processing components. The process begins with the Kinect Azure capturing an IR stream and an RGB stream. The IR stream is processed by the Halo Revealer module, which generates a Depth Map, applies Background Subtraction to isolate the hand, and uses Pixel Enhancement to highlight the fingertip. Simultaneously, the RGB stream is processed by the Hand Tracker module, blending with the IR stream to produce an IR + RGB Blended image, which is then used for hand landmark detection, resulting in a Hand Tracked Image. The detected hand landmarks are passed to the Halo Signal Extractor module to isolate signals related to the fingertip's proximity to the surface. These signals are refined by the Signal Corrector module, which adjusts for noise and inaccuracies, as shown by a heatmap of predicted halo values at a given angle. The corrected signals are then analyzed by the Touch Detector module, which identifies touch events by detecting spikes in the signal patterns over time. The detected touch events are presented as the final output in the Touch Events block, demonstrating a robust system for touch detection using a depth camera}
  \label{fig:system_pipeline}
\end{figure*} 
\subsection{Software}

\textbf{HaloTouch} software is implemented using Python in a Windows Anaconda environment. \textbf{HaloTouch} uses the pyk4a python wrapper for Azure Kinect SDK to access the depth and RGB streams of the camera. We use the unbinned Narrow Field of View mode to obtain a 30 FPS, 640 x 576 depth stream and a longer operating range (0.5 – 3.86m), with the trade-off of a relatively narrower field of view (75 degrees x 65 degrees). The RGB streams also runs at 30 FPS, and are dynamically warped to match the resolution of the depth stream.

We visualize our system pipeline in Figure \ref{fig:system_pipeline}, and we provide more details in each section below.

\subsubsection{Revealing the Halo Effect} \label{sec:Halo_revealer}
Like other static setup systems, such as WorldKit \cite{WorldKit2013Xiao}, we capture 60 frames of background depth images to model the static environment. We then subtract incoming depth frames from this background model to isolate moving hands from static objects in the scene. In contrast, DIRECT \cite{Xiao2016Direct} employs a five-second rolling window of depth data to dynamically update the background model, which is particularly useful for accurately modeling backgrounds when moving objects are present. However, in our case with a largely static background, a simple background capture is sufficient to achieve the desired isolation with lower complexity. After background subtraction, the moving hand or any new object will have non-zero pixel values.

To visualize the Halo effect, we apply a filter that sets all pixel values greater than a threshold value - in our current system, 8 units - to 255 and all the pixel values smaller than 0 to 127. After background subtraction, all objects in the scene show negative pixel values since they are closer to the camera compared to the background surface. The Halo effect, on the other hand, manifests with positive pixel values, indicating that the sensor interprets these pixels as being further away from the camera relative to the background surface. The filtered image renders hands in color grey and Halo effect in color white.

The Halo threshold value of 8 was empirically determined by testing on pilot users and we used this threshold value throughout our experiments. We recommend using a threshold value between 5-10 depending on the camera placement and surface materials (Figure \ref{fig:halo_signal_strength}). A much higher threshold value would diminish the captured Halo effect signal strength, while a lower value is more susceptible to ambient noise.

\subsubsection{Hand Tracking}
Vision-based methods for real-time hand tracking have become increasingly mature, though they still face challenges in certain situations, such as occlusions and variations in lighting conditions. When virtual content is projected onto surfaces, interacting with this content can cause color overlays on the hands in the captured RGB stream images. Our hand tracking relies on Google's MediaPipe Hand Landmarker \cite{mediapipe2023handlandmarker}, a deep learning-based algorithm that can fail abruptly due to color discrepancies on the user’s hands. While it is possible to train a custom hand-tracking model with a dataset featuring overlaid hands, such a model may not generalize well to a broader range of colors when the projected objects change. To address this, we use a computer vision approach that combines infrared (IR) and RGB image streams in a 0.7:0.3 ratio for image blending. This blended image enhances the color contrast between the hands and the projected virtual content, thereby restoring effective hand-tracking functionality.

\subsubsection{Halo Effect Signal Extraction}
As we warp the color image to be the same position and size as the depth image, we can use hand tracking in the color image to find the fingertip in the depth image. We extract a 30x20 patch around the detected fingertip to perform touch detection. The patch area size is empirically determined to capture the full Halo effect signal around the fingertip, although it may preclude detecting closely-spaced fingertips (a limitation of our implementation). We sum all the non-zero values in this patch to form our initial Halo effect signal value and pass this value to the signal corrector.

\subsubsection{Signal Correction}
\begin{figure*}[h]
  \centering  \includegraphics[width=1.0\linewidth]{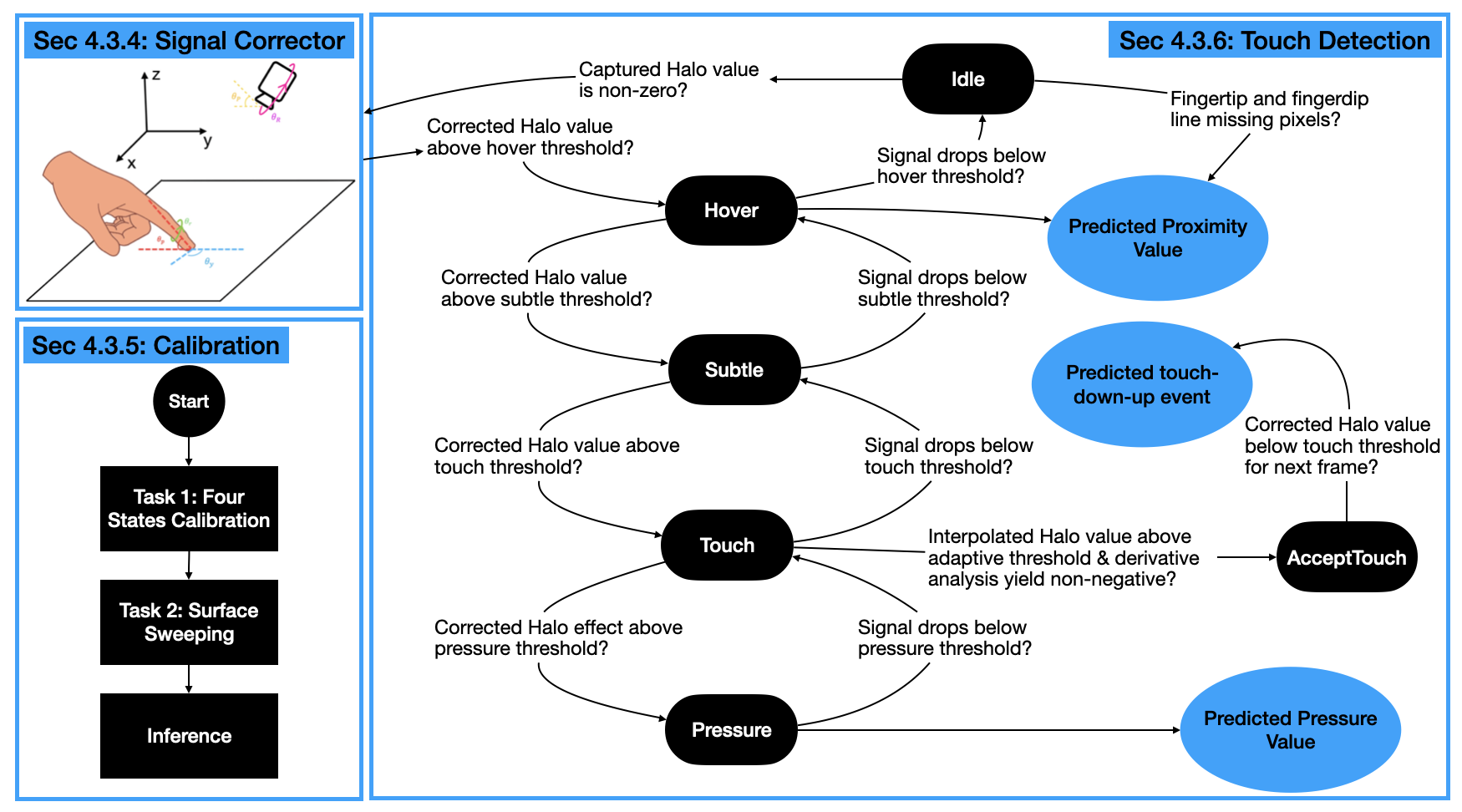}
  \caption{HaloTouch State Machine for Signal Correction, Calibration, and Touch Detection. This diagram outlines the signal processing pipeline for HaloTouch, structured into three key sections: Signal Correction (Sec 4.3.4), Calibration (Sec 4.3.5), and Touch Detection (Sec 4.3.6). The state machine transitions between Idle, Hover, Subtle, Touch, and Pressure states based on corrected halo signal values. The Touch Detection module predicts proximity, touch-down, and pressure values}
  \Description{A flowchart detailing the HaloTouch interaction state machine, divided into three sections: Signal Correction, Calibration, and Touch Detection. The left side highlights Calibration, involving four-state calibration, surface sweeping, and inference. The main diagram presents a state transition model, beginning at Idle and progressing through Hover, Subtle, Touch, and Pressure states based on the corrected halo signal values. The Touch Detection section predicts proximity, touch-down events, and pressure values using an adaptive thresholding and interpolation mechanism. The flowchart visually represents how signal strength determines interaction states in the system.}
  \label{fig:state_machine}
\end{figure*} 
In the context of a touch-down event, many factors could contribute to the Halo effect signal strength in a non-linear manner. Limiting the setup to an empty scene, the Halo effect is influenced by the following factors: the distance between the fingertip and the camera, the 3D orientation of the finger (including roll, pitch, and yaw) relative to the camera, and the amount of finger deformation on the surface. To model these factors, we define four key parameters: x, y, \(\theta_p \) and \(\theta_y \). Our model is based on two key assumptions: (a) users maintain a relatively consistent roll angle \(\theta_r \) while interacting with the surface, and (b) the distance between the camera and the interaction surface remains constant during touch. The parameters x and y represent the fingertip position in the horizontal plane. The pitch angle \(\theta_p \) and yaw angle \(\theta_y \) are derived from the spatial relationship between the fingertip and the fingerdip (distal interphalangeal joint - the first joint closest to the fingertip) using hand landmarks. Note that the system’s robustness is constrained to the 15 to 75 degrees range. Angles below 15 degrees result in a weak Halo effect signal that compromises detection reliability, while angles above 75 degrees can lead to hand-tracking issues caused by fingertip occlusion from the top view.

To minimize user calibration time and reduce the need for extensive datasets, we employ a HistGradientBoosting regressor to predict the strength of the Halo effect and correct the original Halo effect signal value, taking x, y, \(\theta_p \) and \(\theta_y \) as input, and producing a scalar value as output. The regressor model is initially trained using data collected from three pilot users and is fine-tuned for each new user with 20 seconds of calibration data. We use MSE loss, learning rate of 0.1, boosting iterations of 200, max leaf nodes of 31, min samples leaf of 25, and max bins of 255 for regressor training.  Our training data consists of 27000 frames divided into a training-to-validation ratio of 8:2. Three pilot users completed three rounds of sweeping at pitch angles of 15°, 30°, 45°, 60°, and 75° across the surface, as described in Section \ref{sec:calibration}, Task Two. Each round lasted one minute. During each round, the pilot users were instructed to naturally drag their finger across the surface, accompanied by the 3D-printed blocks, to introduce variations to the yaw angle. 

\subsubsection{Calibration}
\label{sec:calibration}
The calibration process consists of two main tasks that users must perform to ensure accurate touch detection.

\begin{figure}[htbp!]
  \centering
  \includegraphics[width=1.0\linewidth]{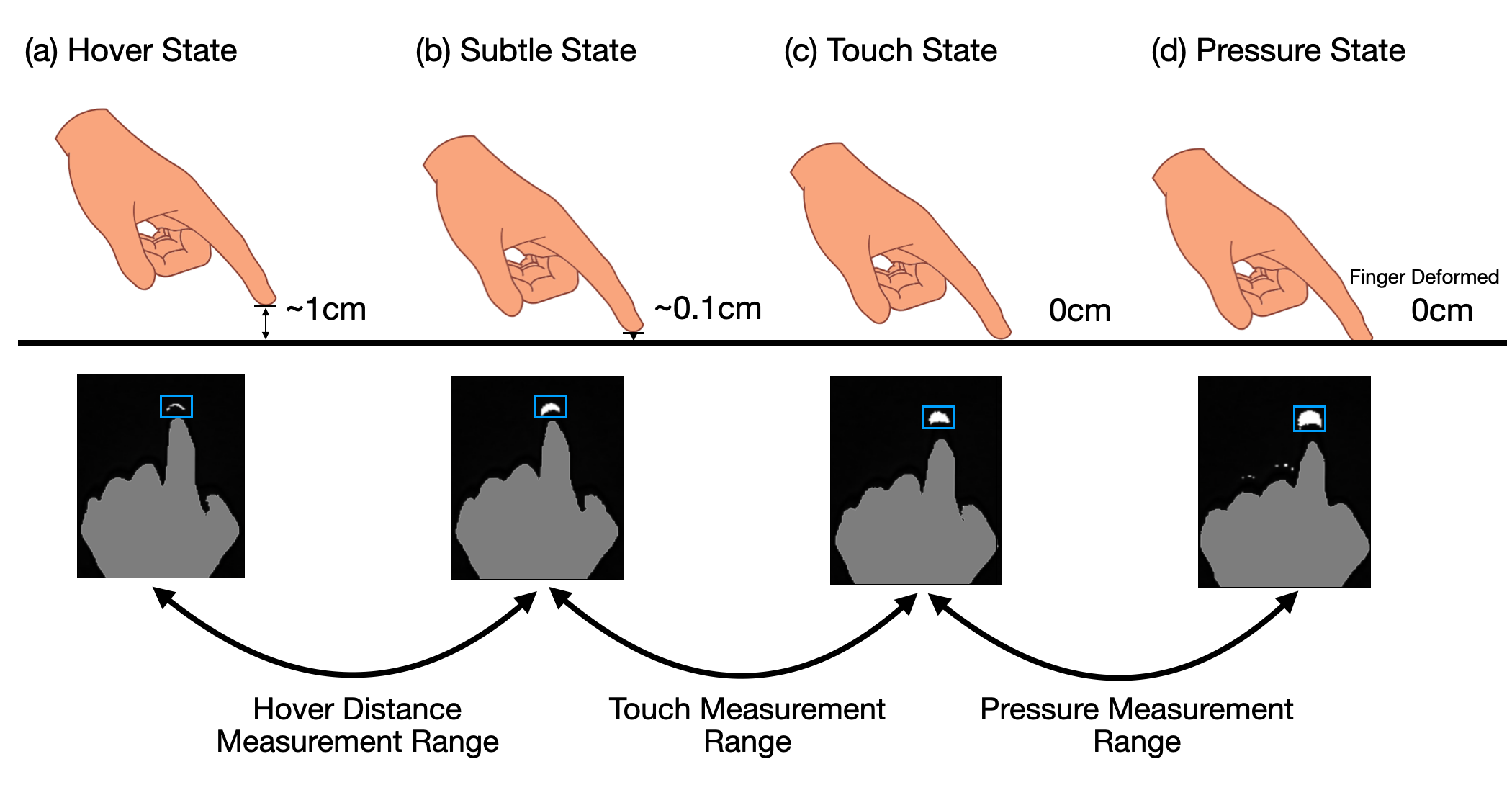}
  \caption{Four calibration states. First row: typical finger to surface distances for each state. Second row: halo effect visualization for each state.}
  \Description{A four-part diagram representing different touch interaction states:(a) Hover State: The fingertip is ~1 cm above the surface, and the depth camera detects a halo effect above the fingertip.(b) Subtle State: The fingertip is very close (~0.1 cm) but not touching, with the halo effect still visible.(c) Touch State: The fingertip contacts the surface (0 cm), with a distinct halo effect present.(d) Pressure State: The fingertip applies force, causing finger deformation while maintaining a strong halo signal.Each state has a corresponding depth visualization, highlighting how the HaloTouch system differentiates interaction states based on halo signal intensity and measurement range.}
  \label{fig:calibration}
\end{figure} 

In the first task, users are required to perform four distinct touch states on the surface to calibrate the system for different levels of interaction:
\begin{enumerate}
    \item Hover State: The finger hovers at a height where the Halo effect just becomes visible. This state captures the initial onset of the Halo effect as the finger gets closer to the surface.
    \item Subtle State: The finger hovers closer to the surface but does not make contact. This state represents a finer level of proximity where the Halo effect becomes more pronounced but still without a physical touch.
    \item Touch State: The finger lightly touches the surface with a relaxed hand posture. This state is the baseline for a standard touch interaction, where the finger makes contact with the surface without exerting additional pressure.
    \item Pressure State: The finger presses down on the surface with a medium amount of force. This state represents a more forceful interaction, which helps in distinguishing between a simple touch and a press action.
\end{enumerate}

In the second task, users quickly sweep their fingers around the touch surface. This sweeping motion allows the system to capture the Halo effect strength at various locations across the surface. To maintain consistency in finger deformation while sweeping, a 3D-printed part angled at 45 degrees is provided for users to rest their index finger on. We employed a fixed 45-degree angle to provide consistency in the collected data. However, in principle, any angle from 15 to 75 degrees could be used, as we trained the signal correction model itself on this range of finger angles.

It is important to note that both calibration tasks are only performed using the dominant hand's index finger. We assume that participants have similar hand structures on both sides; therefore, the calibrated parameters are applied to both hands. However, if significant differences are present, as in the case of a participant with substantially longer artificial nails on the dominant hand compared to the non-dominant hand, separate calibration parameters are set for each hand.

\subsubsection{Touch Detection}

The touch detector identifies touch, proximity, and pressure using the calibrated values from the four calibration states and follows the same interpolation approach described in Equation \ref{equ:linear}, where $x_{min}$ and $x_{max}$ represent the calibrated state values, and $y_{max}$ and $y_{min}$ represent the ground truth maximum and minimum for the corresponding measurement.

\begin{equation}
\label{equ:linear}
y_{measurement} = y_{min} + \frac{x - x_{min}}{x_{max} - x_{min}} (y_{max} - y_{min})
\end{equation}

To detect touch events, the detector interpolates the input Halo effect value x between the subtle state ($x_{min}$) and the touch state ($x_{max}$) values to generate an interpolated value. This value is analyzed using a heuristic-based classifier that relies on thresholding and derivative analysis. The classifier calculates the absolute difference between the current interpolated value and an adaptive baseline. If this difference exceeds a predefined threshold, it proceeds to compute the derivative using the difference between the current and previous interpolated values. A non-negative and relatively large magnitude derivative value confirms a potential touch event by identifying an increasing trend, which is the characteristic of a touch down rather than touch up interaction. The baseline is adaptively updated using exponential smoothing to accommodate signal drift and reduce noise. This approach enables accurate touch detection by leveraging sharp deviations and directional changes to minimize noise and false positives. While the derivative analysis requires two consecutive frames to compute changes accurately, the resulting one-frame delay is a small trade-off for the significant improvement in detection accuracy and reliability.

For typing, a logical state machine debounces touch events, only classifying each touch down and up sequence as a single key input. It verifies valid sequences by requiring touch detection to end with another no-touch detection, reducing false positives during rapid interactions. Although it adds another frame of delay, this approach ensures robust detection, avoiding issues with dropped frames and systematic delays.

The system predicts hover distance by interpolating between the hover and subtle state values using Equation \ref{equ:linear}. Distances just beyond the hover state (10–15~mm) are predicted by interpolating how many pixels still remain on the line between the fingertip and the fingerdip. Similarly, pressure is predicted by interpolating between touch and pressure state values, enabling differentiation between light and firm touch interactions.

\subsection{Camera Placement Impact on Signal Strength}

\begin{figure}[htbp!]
  \centering
  \includegraphics[width=1.0\linewidth]{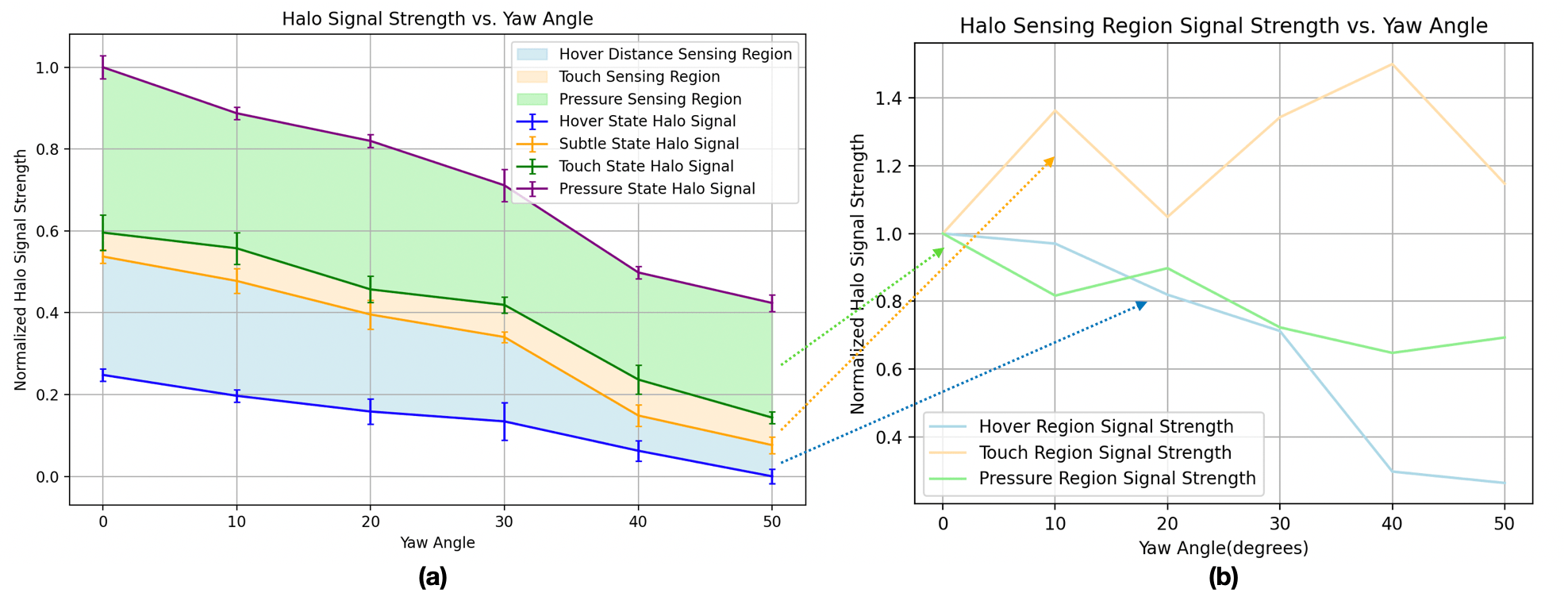}
  \caption{Impact of Camera Yaw Angle on Signal Strength: (a) The Halo signal strength exhibits a similar pattern across all four calibration states when the camera yaw angle ranges from 0 to 50 degrees. (b) Sensing region signal strength: Each sensing region height was calculated and normalized to its 0 degree value}
  \Description{The figure shows two plots illustrating the relationship between Halo signal strength and yaw angle across different interaction states and sensing regions. The left plot presents normalized signal strength for hover, touch, and pressure states, with error bars and shaded regions representing the sensing ranges for hover distance (blue), touch (orange), and pressure (green). Signal strength decreases as the yaw angle increases from 0° to 50°. The right plot highlights changes in normalized signal strength trends for hover, touch, and pressure sensing regions, showing how yaw angle impacts each region differently. Together, the plots demonstrate the effects of yaw angle on signal strength and sensing performance}
  \label{fig:signal_strength_yaw}
\end{figure}

Camera placement affects the strength of the Halo effect signal because changes in yaw angle, pitch angle, and vertical (z) distance alter the intensity and angle of the IR illumination on the surface. We exclude roll angle from consideration because the camera emits a symmetrically distributed IR cone, making roll variations insignificant for signal strength. Figure \ref{fig:signal_strength_yaw} illustrates how yaw angle influences the Halo effect signal strength based on three trials collected at each incremental camera position with the finger placed at the center of the frame. The impact of pitch angle and vertical (z) distance is presented in Appendix \ref{sec:halo_signal_strength}.

It is important to note that system performance depends on the relative differences between calibration states rather than on absolute signal strength. Additional discussion of signal strength and its implications for sensing flexibility can be found in Section \ref{sec:sensing_range_calibration}.

\subsection{Working under Extreme Conditions}
The Halo effect also provides reliable signals in challenging conditions, such as on watery surfaces. Near-infrared (NIR) light, used by IR depth cameras, typically ranges from 850 nm to 940 nm and can penetrate water to a depth of a few millimeters to centimeters. This limited penetration depth is sufficient because typical water blobs on a surface usually have a height of just a few millimeters. We tested our typing application with our pilot users under various water conditions, ranging from multiple small droplets to a thin puddle on the touch surface. No significant performance differences were observed across these conditions, indicating the potential usage of \textbf{HaloTouch} under wet surface scenarios.

\section{Evaluation}
\subsection{Participants and Setup}

We recruited 13 participants (6 male and 7 female) from the local university after obtaining approval from our institution's ethics review board (ID: H19-02782-A010). The data from one participant was excluded from the analysis due to non-compliance with the study protocol during the typing task, and
we report findings based on the remaining 12 valid participants. The 12 participants have an average age of 25.2 (SD=5.1) with skin types ranging from II to VI on the Fitzpatrick scale \cite{fitzpatrick1975soleil}. 9 of the 12 people were dominant in their right hand. All reported they use mobile devices to type regularly. 11 out of 12 people reported that they had used AR/VR devices like Vision Pro, HoloLens, or Quest in the past.

\begin{figure}[htbp!]
  \centering
  \includegraphics[width=1.0\linewidth]{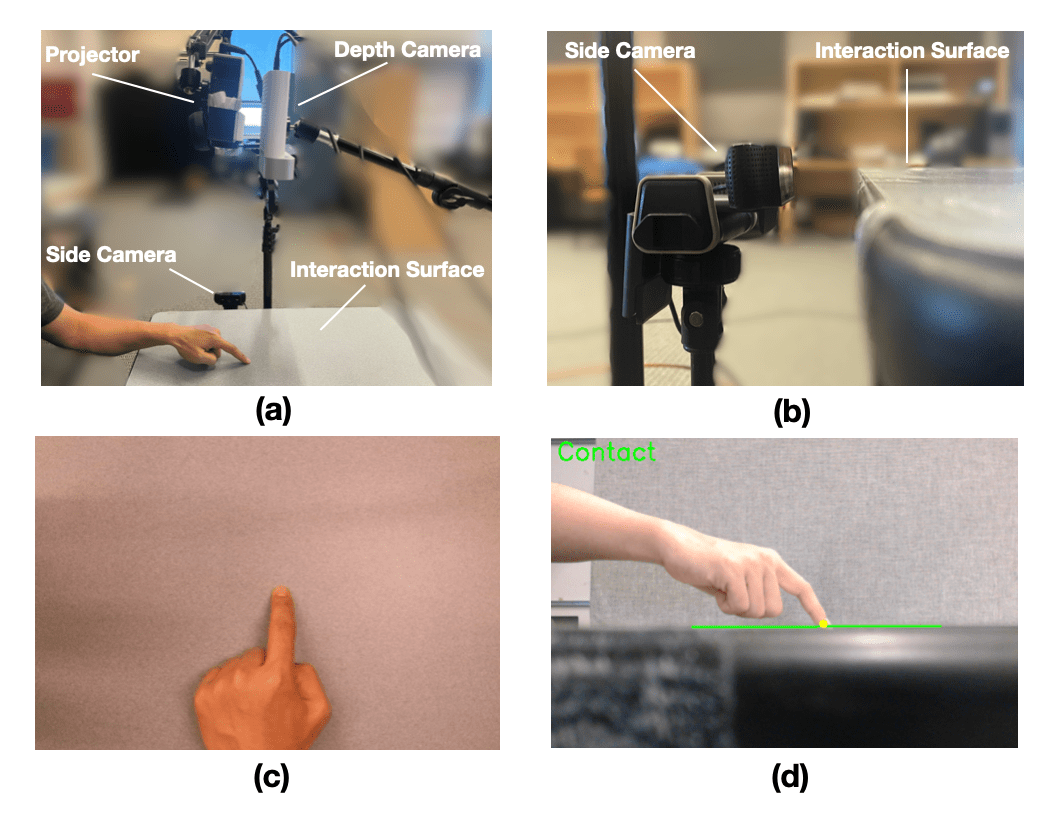}
  \caption{Study Scene Setup: (a) Full setup overview (b) Side-camera aligned with the interaction surface height (c) depth camera RGB view (d) side-camera view}
  \Description{The figure shows four subfigures (a), (b), (c), and (d) that illustrate the setup and perspectives of a touch detection system using a projector, a depth camera, a side camera, and an interaction surface. Subfigure (a) provides a top-down view of the setup, with a projector and a depth camera mounted above the interaction surface and a side camera positioned to capture side views; a hand is shown pointing at the surface to demonstrate touch interaction. Subfigure (b) offers a close-up side view of the side camera, positioned level with the interaction surface, used to capture ground truth data for touch events from a side perspective. Subfigure (c) presents an overhead view of a hand touching the interaction surface with the index finger extended, focusing on the point of contact. Subfigure (d) displays a side view of a hand nearing contact with the interaction surface, with a green line marking the surface level and a label "Contact" indicating the exact moment of touch. These subfigures collectively provide an overview of the experimental setup and the use of multiple camera perspectives to capture accurate touch events from different angles.}
  \label{fig:setup_scene}
\end{figure}

As shown in Figure \ref{fig:setup_scene}, we mounted our Kinect Azure depth camera 0.5~m above the table using extension arms. The projector is mounted right next to the Kinect Azure depth camera to provide projections on the table. An additional side camera is positioned at table height to capture ground truth data for all touch events from a side view. We manually defined a line in the side camera's field of view to align with the surface level, and used Mediapipe's hand tracking to identify fingertip positions. A ground-truth touch event is detected when the fingertip collides with this defined line, enabling robust touch detection and hover distance detection with accuracy of up to 1.2~mm. 

The ground truth data accuracy accounts for hand-tracking jitter. The camera's precision is 0.54~mm per pixel, determined by measuring line lengths in the camera view. Jitter analysis, based on the standard deviation of mean positions over 5 seconds, showed an average of 1.26 pixels, corresponding to a 0.68~mm impact on ground truth accuracy.

Throughout the experiment, participants were instructed to extend their index finger while keeping others near the palm, but could adjust clenching, wrist position, and finger posture. Extending multiple fingers, such as the middle or thumb, affected the Halo effect due to interference. While it is possible to model the dynamic interactions between multiple fingers, this study focuses on single-finger interactions for simplicity.

\subsection{User Study Protocol}
\subsubsection{Task One: Target Pointing}
Sixteen target points were arranged in a 4×4 grid, evenly spaced within a 20×20~cm square area. One point would appear at a time within this area, and participants were instructed to touch it with a 1 second pause between each point. If a touch was registered within a 2.5~cm radius of the target point, the point would be recorded, disappear, and the next point would appear. If no touch was detected, or if the touch occurred outside the 2.5~cm radius, the current point would remain visible. In such cases, participants were asked to retry touching the point with a different finger angle or force. Each participant completed two sets of regular touches (with touch durations between 0.3s and 0.6s) and two sets of swift touches (with touch durations between 0.1s and 0.3s) across five different materials (paper, wood, leather sleeve (suede), plastic, and foam). The points were presented in a random order. Prior to the main test round, participants practiced one set of regular touches and one set of swift touches to get familiar with the protocol, which we do not include in our analysis.

\subsubsection{Task Two: Proximity and Pressure Sensing}

We 3D printed small cylinders of varying heights: 5~mm, 7~mm, 10~mm, 12~mm, and 15~mm. Participants were provided with one cylinder at a time and instructed to place it under their index finger in a relaxed posture. The hover distance was then recorded from the system. Each participant performed 5 heights × 2 repetitions: 10 trials in total.

An iPhone 6 was used in the setup to provide ground truth (0 to 385~g) for pressure force measurements using its 3D touch feature. As infrared light is directly reflected off phone screens, preventing any multipath interference, a layer of regular paper was taped over half of the iPhone 6 screen to enable valid readings by our system. The other half of the screen displayed a normalized force reading (0 to 1). Participants were asked to touch the paper-covered half of the screen and adjust their force to five different levels (0.1, 0.25, 0.5, 0.75, 1) according to the display on the iPhone 6. The output value from our system was recorded for each level. Each participant performed 5 pressure levels × 2 repetitions = 10 trials.

\subsubsection{Task Three: Virtual Keyboard Typing}
\begin{figure}[htbp!]
  \centering  \includegraphics[width=1.0\linewidth]{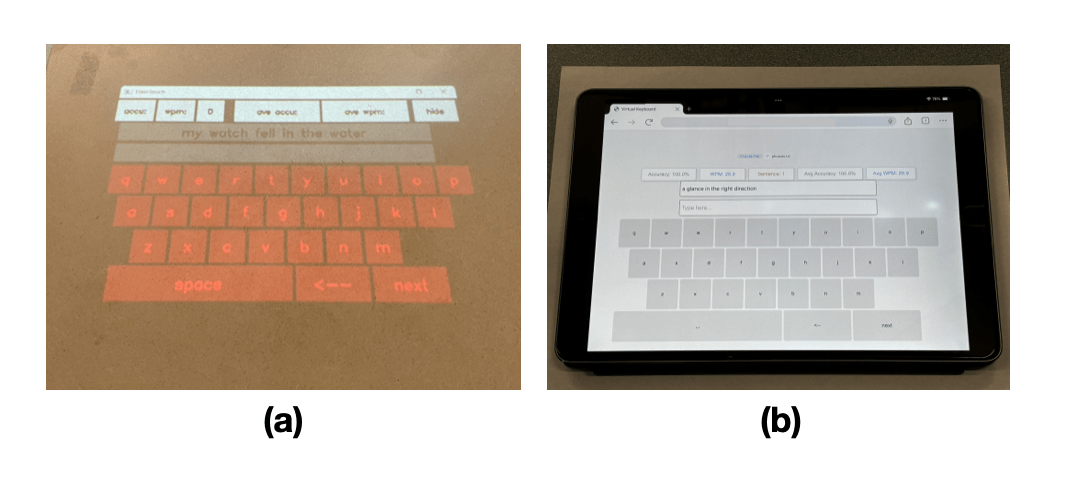}
  \caption{Typing Application: (a) Virtual keyboard in HaloTouch projected on wood surface (b) Mobile keyboard with same layout in IPad}
  \Description{The figure shows two images, (a) and (b), demonstrating different virtual keyboard setups. Image (a) displays a projected keyboard on a flat surface using a projection system, where the keys are red, and a text input area above shows the phrase "my watch fell in the water," along with word suggestions at the top. Image (b) features an iPad on a flat surface displaying a virtual keyboard on its screen with grey keys; the text input box contains the phrase "a glance in the right direction," and performance metrics such as Accuracy, Words Per Minute (WPM), and errors are shown above. Both setups illustrate interactive typing applications designed to evaluate typing performance on various interfaces}
  \label{fig:example_applications}
\end{figure}

A virtual keyboard is presented on the touch surface for the participants (Figure \ref{fig:example_applications}(a)). We used a standard QWERTY keyboard layout, and a standard square key size of 16~mm for our virtual keyboard but with only letters. On top of the keyboard, participants have a reference sentence that they were asked to follow for typing, with their actual input sentence displayed below. The reference sentences were randomly sampled from the widely used Mackenzie phrase set, which contains 500 phrases \cite{MackKenzie2003PhraseSet}. Participants can also view their current sentence level accuracy and words per minute (WPM), so that they can adhere to the different typing goals. We played typing sound feedback at each click for better training and experience, per \cite{Ma2015HapticKF}.

We ran a 10-minute warm up session for each participant after a brief introduction of the keyboard elements. In each test round, we asked the participants to type on the virtual keyboard across three tasks, described as follows:
\begin{itemize}
    \item T1 - Prioritize accuracy: type with more caution to make sure you achieve high accuracy
    \item T2 - Balance accuracy and speed: type fast like  you would on your mobile device, but try to maintain an accuracy of 90\%.
    \item T3 - Prioritize speed: read the reference sentence first, plan out the finger placement trajectory and then type as fast as possible.
\end{itemize}

For each task we asked the participants to type 8 sentences, and we recorded the average typing accuracy and WPM at the end of each task. We ran N=2 repeats for each task. We note while running the test round, we asked the participants not to use the ``delete'' key unless it is a mistake that is due to their own behavior. This protocol yields data that avoids user bias towards higher typing accuracy, better reflecting the user true typing input statistics towards the system.

For comparison, we ran one round of the same 8 sentences typed on an Apple iPad Pro (11 inches 2rd Gen) and on a Microsoft HoloLens 2 for each participant. The keyboard (Figure \ref{fig:example_applications}(b)) follows exactly the same layout and size implemented as a web application. On HoloLens, the keyboard is implemented using crossing buttons, similar to the default HoloLens 2 keyboard.

\subsection{Results}
\subsubsection{Touch Down Accuracy}
\begin{figure}[htbp!]
  \centering
  \includegraphics[width=1.0\linewidth]{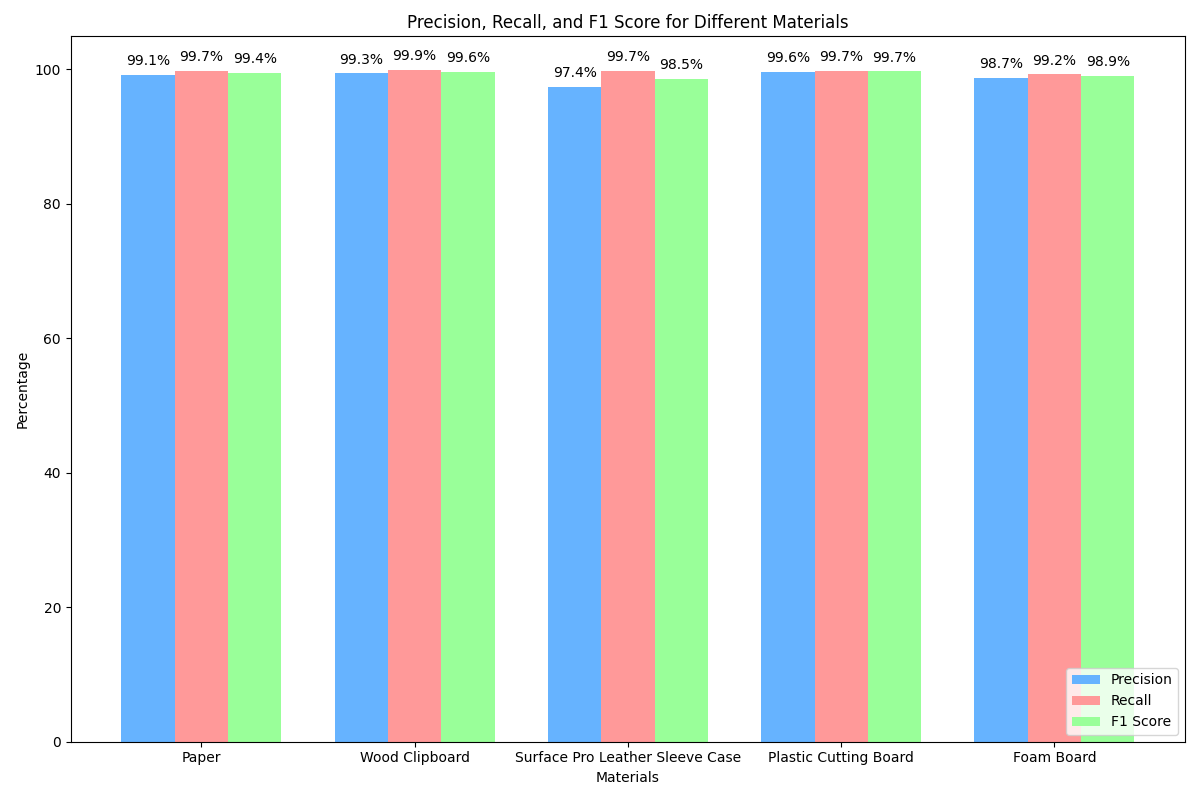}
  \caption{Precision, Recall, and F1 Score for touch detection across five materials: paper, wood, leather sleeve (suede), plastic, and foam in our study.}
  \Description{The bar chart presents the Precision, Recall, and F1 Score percentages for touch detection accuracy across five different materials: Paper, Wood Clipboard, Surface Pro Leather Sleeve Case, Plastic Cutting Board, and Foam Board. The x-axis represents the materials, while the y-axis shows the percentage values from 0 to 100\%. Each material has three color-coded bars indicating Precision (blue), Recall (red), and F1 Score (green). The chart demonstrates that the touch detection system achieves high accuracy across different materials, with all metrics—Precision, Recall, and F1 Score—consistently above 98\% for most surfaces, highlighting the system's reliability and effectiveness regardless of the material used}
  \label{fig:touch_accuracy_all_materials}
\end{figure} 
Across 12 participants, we obtained 3840 test touch points. Of these, 13 reported no touch contact (0.34\%), 45 reported more than one touch point (1.17\%). We report an average precision of 98.8\%, recall of 99.7\% and F1 score of 99.2\% among all the data points. We show the precision, recall, and F1 score for each material in  Figure \ref{fig:touch_accuracy_all_materials}.
We see no significant difference (p>0.01, ANOVA with Tukey HSD) on the false negative touch counts among material pairs. However, we see significant difference (p<0.01, ANOVA with Tukey HSD) on the false positive counts between suede and plastic, and significant difference among other material pairs. We see no significant difference (p>0.1, ANOVA) on precision between regular touches and swift touches. Outside of the 3840 points, our system detected 31 points which are outside of the reference point radius. 

\subsubsection{Spatial Accuracy}

\begin{figure}[htbp!]
  \centering
  \includegraphics[width=0.8\linewidth]{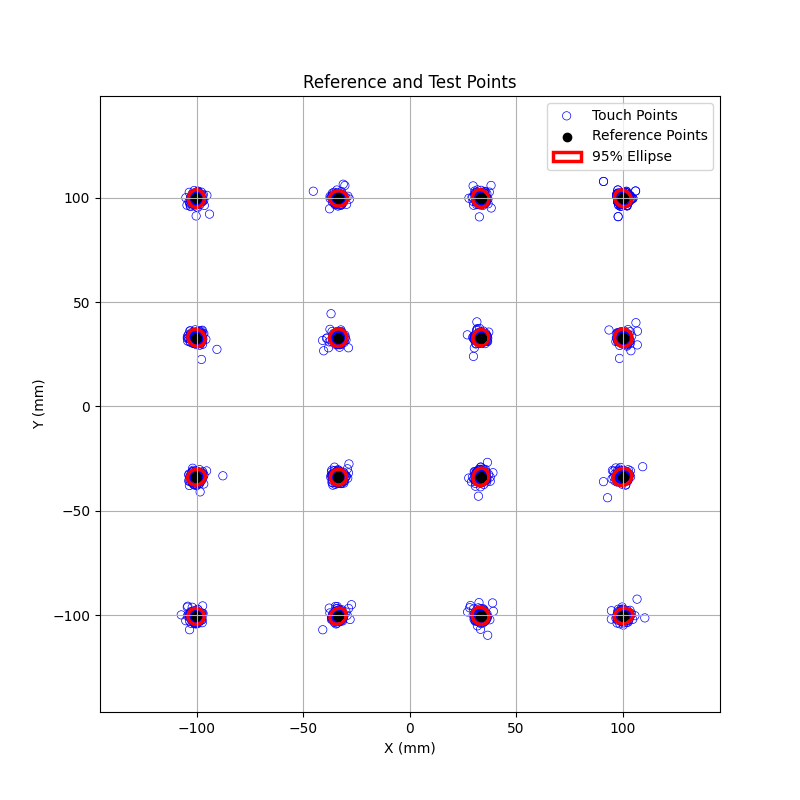}
  \caption{Scatter-plot of all touch points in task one, plotted with 95\% confidence ellipses.}
  \Description{The scatter plot illustrates the distribution of reference and test touch points on a 2D plane to assess touch input detection accuracy, with both the x-axis and y-axis representing coordinates in millimeters (mm) from -150 mm to 150 mm. The plot features several clusters centered around black dots, which represent the Reference Points where touch inputs were expected. Surrounding each reference point are blue open circles indicating the Touch Points, or the actual detected touch locations during testing. Each cluster is enclosed by a red 95\% Ellipse, showing the area within which 95\% of the detected touch points are expected to fall, providing a visual measure of the variability and spread around each reference point. This plot demonstrates how closely the detected touch points align with their reference points, highlighting the accuracy and precision of the touch detection system. The tight clustering of points and the size of the ellipses offer a clear visual representation of the system's performance}
  \label{fig:scatterplot_touch_points}
\end{figure}

We followed the same analysis procedure as MRTouch \cite{Xiao2016Direct}. We experienced a similar systematic offset of all received touch points, which is 7~mm to the left. The shift is constant across all materials and users. We subtracted the global average offset from all touch points for the following analysis. We also removed 61 outlier points (1.6\%) which lay more than three standard deviations from the reference point. These outliers primarily resulted from user errors and inaccuracies. For instance, some users sped up the point-clicking process after becoming familiar with the task, occasionally failing to lift their fingers fully before clicking the next point. This behavior introduced inconsistencies, which were removed to maintain the integrity and accuracy of the dataset. Across all materials and all participants, we report a global mean Euclidean error of 5.5~mm (SD = 0.69).

We see no significant difference between the regular touch and the swift touch conditions (p>0.01, ANOVA). Pair-wise Tukey HSD also shows no significant difference (p>0.01) among the 5 materials. 
We plot the 95\% confidence ellipses for all the received points in Figure \ref{fig:scatterplot_touch_points}. On average, a 16~mm diameter button would capture 95\% of touches, which provides additional support for the size selection in our virtual keyboard implementation.

\subsubsection{Latency}
Our pipeline has an overall mean motion-to-graphics latency of 150~ms, measured using an external high FPS camera. This matches well with the theoretical latency of 146~ms to 186~ms, composed primarily of camera buffer image loading (33~ms), media pipeline hand tracking (55\texttildenew 85~ms), the rest of the processing pipeline(25\texttildenew 35~ms), and projector delay (33~ms).

\subsubsection{Touch Point Threshold}
We measure our touch point threshold by averaging all the vertical travel distance of detected fingertips from side camera for each valid key input. We make the assumption that for each key input, the distance of finger traveling down to touch the key and traveling up before aiming for the next key is the same. Our system has a touch point threshold of 4.97~mm, averaging across all participants in the typing application.

\subsubsection{Proximity Accuracy}
\begin{figure}[htbp!]
  \centering
  \includegraphics[width=0.8\linewidth]{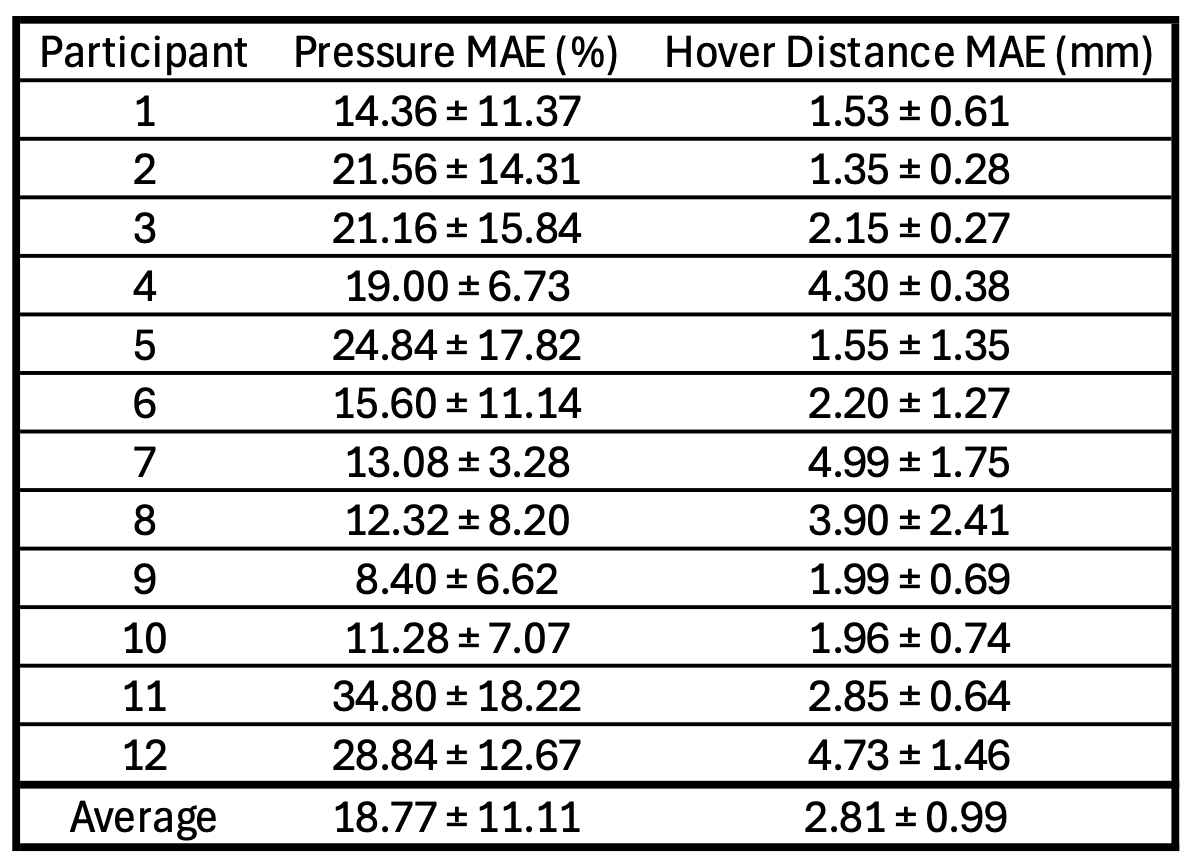}
  \caption{Participants pressure MAE and hover distance MAE breakdown}
  \Description{The table presents the Mean Absolute Error (MAE) for pressure and hover distance measurements for 12 participants in a user study, showing the accuracy and variability of these measurements. The Pressure MAE (\%) column lists the error in pressure detection as a percentage with its standard deviation, while the Hover Distance MAE (mm) column provides the error in detecting hover distance in millimeters along with the standard deviation. The values for Pressure MAE range from 8.40 ± 6.62\% (Participant 9) to 34.80 ± 18.22\% (Participant 11), and the Hover Distance MAE values range from 1.35 ± 0.28 mm (Participant 2) to 4.99 ± 1.75 mm (Participant 7). The average Pressure MAE across all participants is 18.77 ± 11.11\%, and the average Hover Distance MAE is 1.53 ± 0.61 mm. This table provides a comprehensive overview of the error rates in pressure and hover distance detection for each participant, revealing the variability in system performance across different users}
  \label{fig:pressure_hover_MAE}
\end{figure}

We show the overall results per participant in Figure \ref{fig:pressure_hover_MAE}. We report an average hover distance error of 2.81 ± 0.99~mm. Kruskal-Wallis tests indicated no significant differences between the distance targets (H = 5.42, df = 5, p > 0.001), but indicated significant differences between participants (H = 34.28, df = 12, p <0.001).

\subsubsection{Pressure Accuracy}

We show the overall results per participant in Figure \ref{fig:pressure_hover_MAE}. We report an average pressure error of 18.77 ± 11.11\%. A Kruskal-Wallis test indicated no significant differences between the pressure targets (H = 8.69, df = 5, p > 0.001), and also indicated no significant differences between participants (H = 28.61, df = 12, p >0.001).

\subsubsection{Typing Performance}\label{sec:typing_performance}
\begin{figure}[htbp!]
  \centering
  \includegraphics[width=0.9\linewidth]{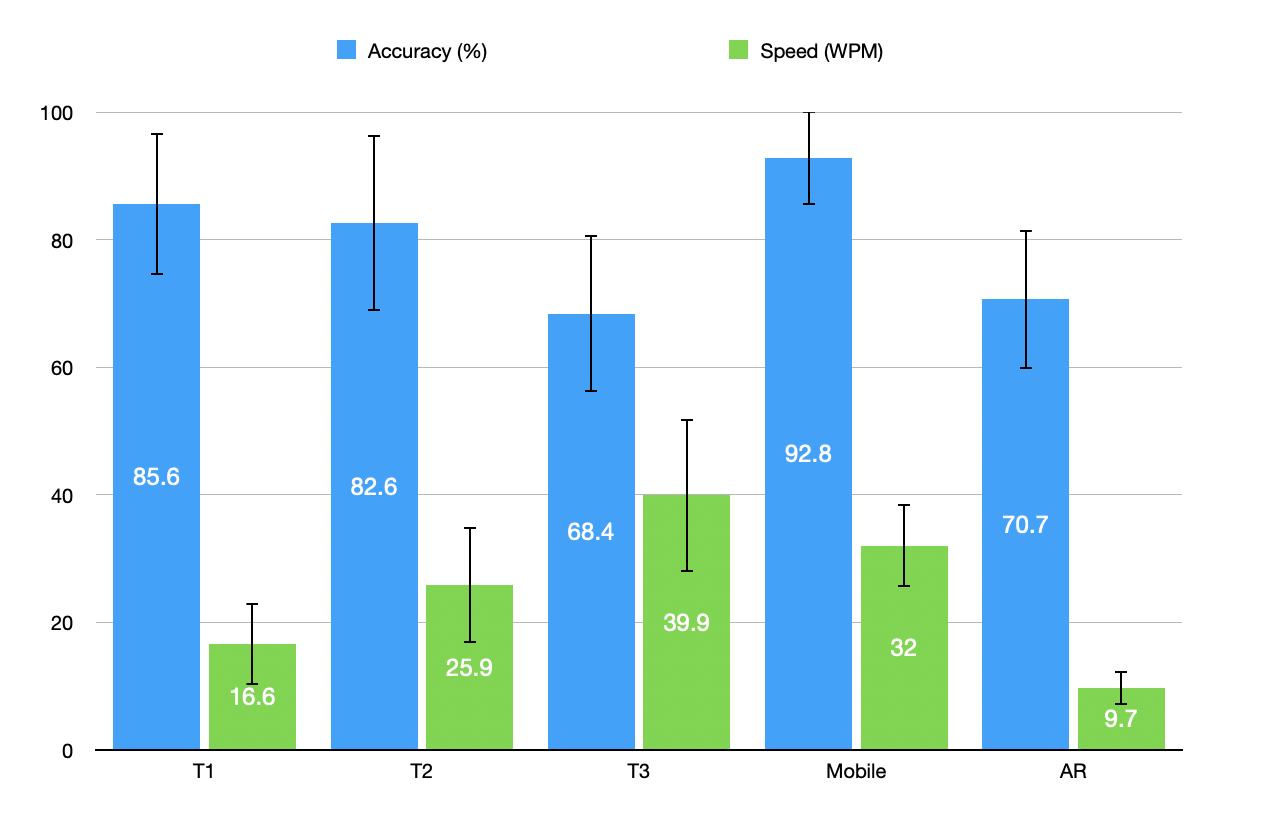}
  \caption{Typing accuracy and speed (WPM) under five conditions: T1, T2, T3, Mobile, and AR}
  \Description{The bar chart compares typing accuracy and speed, measured in words per minute (WPM), across five conditions: T1, T2, T3, Mobile, and AR. The x-axis represents these conditions, while the y-axis shows percentages for accuracy and speed in WPM, ranging from 0 to 100. Each condition has a blue bar for Accuracy (\%) and a green bar for Speed (WPM), with error bars indicating variability.}
  \label{fig:typing_accuracy_wpm}
\end{figure} 

\begin{figure}[htbp!]
  \centering
  \includegraphics[width=0.9\linewidth]{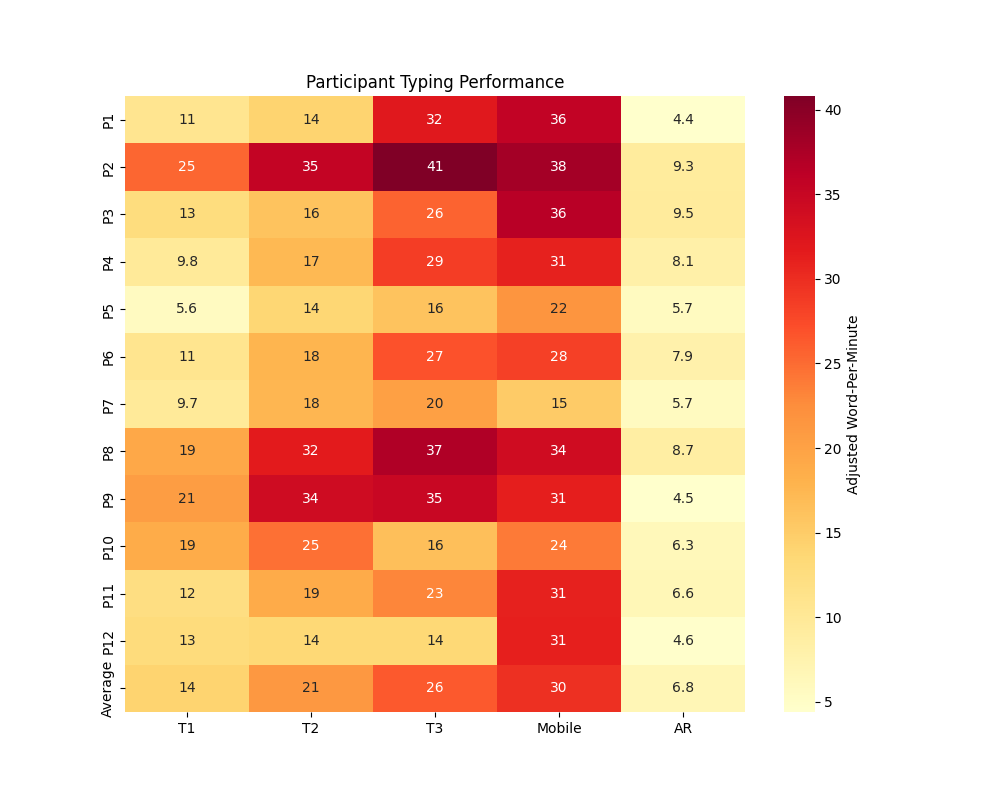}
  \caption{Typing speeds (AWPM) for 12 participants across five conditions. The color gradient visualizes performance differences, with Mobile generally yielding higher speeds and AR having the lowest}
  \Description{The heatmap shows the adjusted typing speed in words per minute (WPM) for 12 participants (P1 to P12) across five conditions: T1, T2, T3, Mobile, and AR. The x-axis represents these conditions, and the y-axis lists the participants along with the average typing performance across all participants. The color gradient, ranging from light yellow to dark red, represents the WPM values, where lighter colors indicate lower speeds and darker colors indicate higher speeds.}
  \label{fig:typing_awpm}
\end{figure} 

To evaluate typing performance, we measured both text entry speed and accuracy. Text entry rate (in WPM) was calculated by taking the time difference between the first and last keystrokes for each phrase. For sentence-level accuracy, we split each input sentence into words and compared them to their corresponding reference words at the character level, then reported the average accuracy per participant. We show the task-wise typing accuracy and speed comparison along with the AR and mobile device typing performance in Figure \ref{fig:typing_accuracy_wpm}.

To present a more intuitive measure combining typing speed and accuracy, we calculated Adjusted Words Per Minute (AWPM), where AWPM = WPM × accuracy rate. Unlike raw WPM, which only considers the number of words typed in a minute, AWPM considers the error rate as well, providing a more accurate reflection of typing proficiency. We show the AWPM for each user in Figure \ref{fig:typing_awpm}. Additional metrics for text entry including Uncorrected Error Rate (UER) and Corrected Error Rate (CER) (using definition by \cite{Zhang2019ByondInput}) can be found in Appendix \ref{sec:typing_error}.
Using the ground truth data provided from the side camera, we calculated the fingertip-to-surface hover distance when no touch is detected during each typing round for each participant, and produced average participant hover distances for each typing task (T1, T2, T3). We perform Pearson correlation test between AWPM and average hover distance among all participants and found a strong positive correlation (Pearson correlation coefficient = 0.715, p-value < 0.01) for T3, a strong positive correlation for T2 (Pearson correlation coefficient = 0.698, p-value < 0.05), and an insignificantly negative correlation for T1 (Pearson correlation coefficient = -0.197, p-value > 0.05).

\section{Example Applications}

\subsection{Gesture Augmented Virtual Drumming}
The HaloTouch system's proximity detection capability allows it to accurately sense the hover distance of the user's fingertip above a surface. This feature is utilized in our virtual drumming app, where users can engage in an expressive drumming experience that goes beyond simple tapping. The app leverages both hands of the user to create a dynamic and rich interactive musical environment.

With one hand, users can play the virtual drum by tapping on any surface, producing realistic drum sounds. Meanwhile, the other hand can be used to hover at various heights above the surface. The distance of this hovering hand from the surface is continuously detected by \textbf{HaloTouch}, allowing it to modulate the characteristics of the drum sound in real time. For instance, hovering the hand closer to the surface can enhance the lower frequencies of the drum sound, creating a deeper, bass-like tone, while hovering higher could emphasize higher frequencies, generating sharper, snappier sounds. This modulation effectively alters the harmonic content of the drum sound, creating a wide range of expressive possibilities.

\subsection{Pressure Sensitive Painting}
The \textbf{HaloTouch} system's pressure detection capability enables it to sense the amount of force a user applies on a surface. This feature is utilized in our painting application to provide a more natural and dynamic drawing experience by allowing users to modulate the thickness and style of brush strokes based on varying pressure levels, akin to using traditional art tools like brushes, pens, or calligraphy brushes.

In this application, the amount of pressure detected by \textbf{HaloTouch} is seamlessly mapped to stroke attributes such as width, opacity, and texture. Light pressure results in fine, delicate lines suitable for detailing and sketching, while heavier pressure creates broader, more pronounced strokes ideal for bold expressions or filling large areas. This enables users to achieve natural calligraphy effects and provides more freedom and flexibility compared to traditional fixed-stroke painting tools.

The app enhances the painting experience by allowing for smooth transitions between different stroke types within a single brush, enabling artists to create varied textures and effects without having to switch tools frequently. For example, an artist can start with a light stroke to outline a shape and then gradually increase the pressure to add depth, shading, or emphasis—all within the same continuous motion. This mimics the fluidity of traditional painting techniques, where artists control their brush or pen strokes by applying varying amounts of pressure.
\begin{figure}[htbp!]
  \centering
  \includegraphics[width=1.0\linewidth]{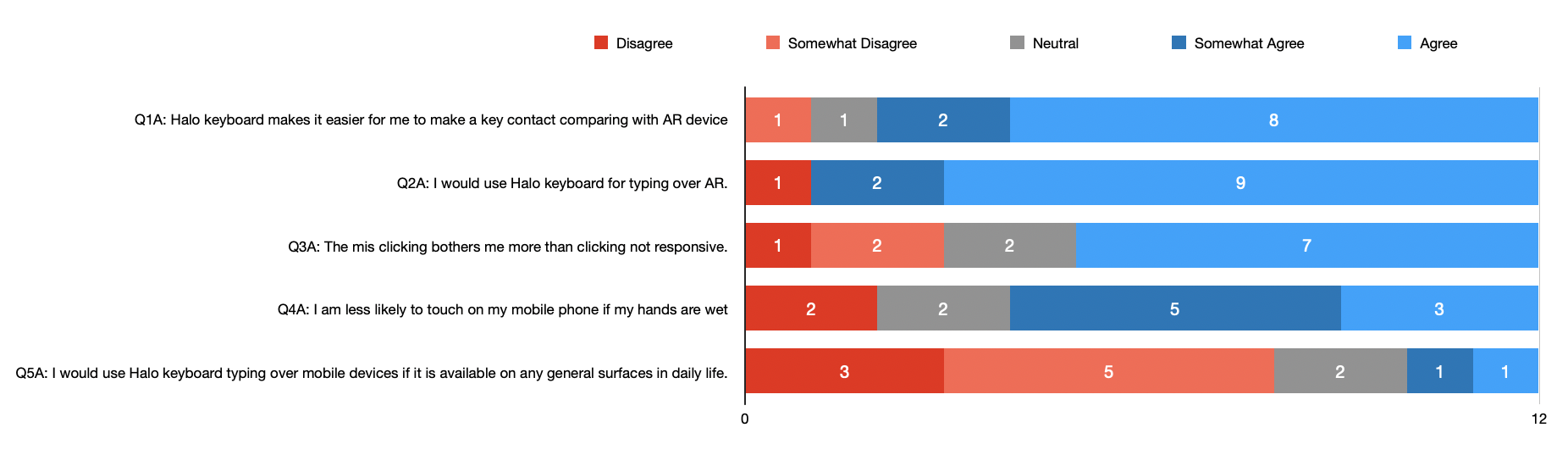}
  \caption{Results from the 5-point Likert scale questions}
  \Description{The stacked bar chart shows responses from a questionnaire comparing user preferences and experiences with the Halo keyboard versus an AR device and mobile devices, featuring five questions (Q1A to Q5A) with responses categorized as Disagree, Somewhat Disagree, Neutral, Somewhat Agree, and Agree. The chart indicates a general preference for the Halo keyboard over AR for typing, while opinions are more divided when comparing it to mobile devices, reflecting varied user preferences and concerns.}
  \label{fig:questionnaire}
\end{figure} 

\section{Discussion and Future Work}

\subsection{Technical Metrics In Comparison with Other Systems}
\label{sec:tech_metrics}
\textbf{HaloTouch} is a comprehensive sensing system that enhances touch interaction on general surfaces by achieving high touch-down accuracy, spatial accuracy, touch point threshold and no instrumentation. In this section, we discuss the key advantages of \textbf{HaloTouch} compared to existing systems, focusing on its touch point threshold, versatility, and ability to support diverse input interactions.

Firstly, \textbf{HaloTouch} achieves high touch-down accuracy (>99.0\%), comparable to systems like Electrick{\cite{Zhang2017Electrick}}, ShadowTouch{\cite{Liang2023ShadowTouch}}, and Direct{\cite{Xiao2016Direct}}, and provides high spatial accuracy (<10~mm), similar to Electrick{\cite{Zhang2017Electrick}}, Direct{\cite{Xiao2016Direct}}, and FarOut{\cite{Shen2021Farout}}. While \textbf{HaloTouch} operates with slightly higher latency than some other systems, our typing application demonstrates that a latency of 150~ms does not noticeably impact user experience or performance.

To further enable rapid and accurate touch input, minimizing the touch point threshold is critical. \textbf{HaloTouch} achieves a remarkably low touch point threshold of 4.97~mm, outperforming other vision-based approaches such as TapLight{\cite{Streli2023LightSpeckle}}, Direct{\cite{Xiao2016Direct}}, FarOut{\cite{Shen2021Farout}}, and Tripad{\cite{Dupre2024Tripad}}. We explain the causes of false positives and false negatives in virtual keyboard typing by discussing the concept of the touch point threshold. Imagine that touching a virtual key is like pressing a physical button with a certain ``height'' that must be fully pressed down to register a valid touch. The touch point thresholds defines this virtual height. A higher touch point thresholds requires more effort from users to transition from one button to another. In a virtual keyboard, however, without the tactile feedback of physical keys, users often cross into adjacent keys without resetting to the required touch point thresholds. This misalignment is a major cause of false positives or false negatives, depending on the specific threshold settings. Capacitive touch screens, for instance, effectively have a zero touch point thresholds, allowing users to transition smoothly between keys without restrictions. The positive correlation between touch point thresholds and typing performance in T2 and T3 also indicates that users who are used to lift fingers up higher while typing are able to accommodate for systems with a higher touch point threshold. Although systems like DIRECT \cite{Xiao2016Direct}, TapLight \cite{Streli2023LightSpeckle}, and Tripad \cite{Dupre2024Tripad} excel in touch-down and spatial accuracy, their high touch point thresholds can lead to a high false positive rate during fast interactions. In contrast, while maintaining high touch-down accuracy and spatial accuracy, \textbf{HaloTouch}’s low touch point thresholds significantly enhances input speed and accuracy, particularly in rapid typing scenarios.

Another key advantage of \textbf{HaloTouch} is its ability to operate on most general materials without requiring surface instrumentation or additional sensors attached to the user. These features are vital for achieving true ad hoc sensing without limiting the user's freedom. In contrast, other systems with high material compatibility and low instrumentation requirements often compromise on spatial accuracy (e.g., Shi et al. \cite{Shi2020FingerIMU}) or have a higher touch point threshold (e.g., Tripad \cite{Dupre2024Tripad}), making rapid and accurate touch interactions difficult to support.

Moreover, \textbf{HaloTouch}'s capability to detect pressure and proximity enables a richer input interaction space. This includes in-air gestures, pressure-modulated inputs as demonstrated in our example applications, and the combination of hover, touch, and pressure—all using only a commercially available depth camera. With these capabilities, \textbf{HaloTouch} extends the interaction modalities on general surfaces, offering a versatile and user-friendly solution for touch-based applications.

\subsection{Learning Effect \& User Experience}
We observed a strong learning effect in the keyboard typing study, which can be explained by two main causes. First, most users are accustomed to typing with their thumbs on mobile devices. While some users reported using their index finger for typing on larger devices like iPads, they do so less frequently. Given that typing performance in T3 for the Halo keyboard was nearly on par with that on mobile devices, it suggests that users can adapt to typing with their index finger after some training and practice. Second, typing on flat surfaces lacks the haptic feedback that users rely on. Although we provided sound feedback when a key was clicked, some participants reported difficulty in knowing whether they had successfully touched a key during the warmup round. For instance, P7 mentioned, ``I got really tired in the learning round where I had to constantly make sure if I made contact with the key or not.'' Similarly, P9 suggested, ``I would enjoy the typing experience even more if the keyboard gives me a vibration.''

Users also faced challenges in estimating how far they needed to reach out in the air to make a key click in HoloLens. P5 described the experience as, ``It is hard to click on the key in AR because I cannot feel where the key is relative to my hand.'' Conversely, P2 expressed a preference for AR, stating, ``I really like how I can see my finger cursor is highlighted in AR, and that helps me know when I am clicking a key.'' P12 had mixed feelings: ``The AR keyboard requires me to move my whole arm to type, while the Halo keyboard is easier to click on, but the mis-clicks slow down my typing speed.'' Notably, 10 out of 12 users (P1, P3-P11) felt that the Halo keyboard made it easier to make key contact compared to the AR device (Q1A).

We noticed that users who were more confident tended to type faster, while those who were more cautious typed slower. Specifically, users who experienced more false positives during the warmup round tended to type more cautiously and slowly in the experiment round. In contrast, users who typed comfortably in the warmup round often ended up typing even faster in the experiment round. Our implementation of filtering out artifacts based on finger movement speed favors users who type more quickly.

\subsection{Privacy \& Trust}
Almost all users (11 out of 12) preferred typing with the Halo keyboard over the AR keyboard (Q2A). However, 75\% of them still favored using their mobile phones, even if the Halo keyboard were available on any general surface in daily life (Q5A). P9 highlighted a critical issue regarding privacy: ``Halo keyboard types quite nicely for me, and it would be really cool to have this kind of keyboard everywhere, but I still like my phone better because I don’t want others to see what I am doing.'' This sentiment underscores the ongoing trade-off between privacy and convenience. While we believe that \textbf{HaloTouch} could offer significant convenience if integrated into everyday life, preserving user privacy remains essential. A potential future work to address this concern could involve deploying \textbf{HaloTouch} onto AR headsets. While our system currently works on Kinect Azure, we have confirmed the presence of the Halo effect at varying surfaces and heights (up to 1~m) using HoloLens 2 IR camera feed, validating our working theory. The primary challenge for deployment would involve optimizing for a moving IR light source coupled with a moving plane, ensuring consistent and reliable Halo effect signal strength while maintaining user privacy.

We attribute the lack of correlation between T1 and AWPM to participants' behavior during experiments. In T1, the emphasis on accuracy may have led participants to prioritize precision over trust-building, suppressing natural typing habits. In T2 and T3, relaxed accuracy requirements likely encouraged more natural typing, improving performance as discussed in Section \ref{sec:tech_metrics}.

Beyond technical feasibility, P8 raised another point: ``I prefer my phone because I am more comfortable interacting with my own belongings.'' This touches on the broader issue of trust between humans and computers, which remains an open research area. Although this paper focuses on enabling new interaction techniques, future research could explore creating an environment where users feel comfortable interacting with any passive surface, much like they do with their personal devices today. This raises new research questions: Can people develop trust in an object they already own if it is augmented with input techniques? Can the \textbf{HaloTouch} technique encourage people to trust their environment more over time?

\subsection{Sensing Range and Calibration}
\label{sec:sensing_range_calibration}
Our study setup features a sensing distance of 0.5 meters and no yaw or pitch angle on the camera. With our current setup, we support an interaction area of up to 30 cm × 20 cm with index fingertips at a camera distance of 0.5 meters. Our characterization results in Figure \ref{fig:signal_strength_yaw}, \ref{fig:signal_strength_z}, and \ref{fig:signal_strength_pitch} show that the performance of the touch measurement and pressure measurement is not heavily impacted by distance and angle, whereas the resolution of the proximity can be impaired by the distance and the angle of the camera placement. Particularly, the Halo effect strength for proximity falls below 50\% at distance of 80~cm, yaw angle of 25 degrees, and pitch angle of 15 degrees. the Halo effect strength for touch input is maintained above 75\% up to distance of 100~cm, yaw angle of 50 degrees, and pitch angle of 45 degrees. The Halo effect strength for pressure is maintained above 75\% for up to distance of 70~cm, yaw angle of 35 degrees, and pitch angle of 45 degrees. The Halo effect strength for pressure is amplified between pitch angle -35 degrees to 0 degrees. 

The extent to which the Halo effect becomes too weak to support reliable touch detection also depends on factors such as the properties of the surface material. For example, our study found that foam board generates significantly less Halo effect compared to other materials, while the leather sleeve produces a much stronger Halo effect. 

Despite the comprehensive sensing capabilities that HaloTouch offers, a 20-second calibration is required for each user to accurately capture the finger response across different positions in the interaction area. We additionally characterized the Halo effect given various camera positions. Such characterization can provide extra data to train a more robust algorithm that is free of calibration in the future. 

\section{Conclusion}
We introduced \textbf{HaloTouch}, a novel input sensing technique that leverages multipath interference from commercially available depth cameras to enable high-accuracy touch input on a variety of passive surfaces without the need for additional hardware or surface instrumentation. \textbf{HaloTouch} excels in its ability to provide comprehensive sensing capabilities, combining both pressure and proximity detection with impressive accuracy across multiple surface types. Our exploration of the Halo Effect has demonstrated the potential of this underutilized phenomenon for surpassing noise thresholds in depth sensors, making it a robust solution for real-world applications. The two-part study validated the technical performance and versatility of HaloTouch in supporting rapid, ad hoc interactions in diverse environments, including extreme settings like watery surfaces.

Our work contributes to the growing field of on-world interfaces by providing a flexible, multimodal touch input system that broadens the potential for interaction in uninstrumented environments. Moving forward, \textbf{HaloTouch} paves the way for future research in expanding the capabilities of depth cameras and enhancing the seamless integration of digital interactions with the physical world.

\begin{acks}
This work was supported in part by the Natural Science and Engineering Research Council of Canada (NSERC) under Discovery Grant RGPIN-2019-05624 and by Rogers Communications Inc. under the Rogers-UBC Collaborative Research Grant: Augmented and
Virtual Reality.
\end{acks}

\bibliographystyle{ACM-Reference-Format}
\bibliography{sample-base}


\begin{thebibliography}{71}


\ifx \showCODEN    \undefined \def \showCODEN     #1{\unskip}     \fi
\ifx \showDOI      \undefined \def \showDOI       #1{#1}\fi
\ifx \showISBNx    \undefined \def \showISBNx     #1{\unskip}     \fi
\ifx \showISBNxiii \undefined \def \showISBNxiii  #1{\unskip}     \fi
\ifx \showISSN     \undefined \def \showISSN      #1{\unskip}     \fi
\ifx \showLCCN     \undefined \def \showLCCN      #1{\unskip}     \fi
\ifx \shownote     \undefined \def \shownote      #1{#1}          \fi
\ifx \showarticletitle \undefined \def \showarticletitle #1{#1}   \fi
\ifx \showURL      \undefined \def \showURL       {\relax}        \fi
\providecommand\bibfield[2]{#2}
\providecommand\bibinfo[2]{#2}
\providecommand\natexlab[1]{#1}
\providecommand\showeprint[2][]{arXiv:#2}

\bibitem[Abdlkarim et~al\mbox{.}(2023)]%
        {Abdlkarim2023MetaQuest}
\bibfield{author}{\bibinfo{person}{Diar Abdlkarim}, \bibinfo{person}{Massimiliano Di~Luca}, \bibinfo{person}{Poppy Aves}, \bibinfo{person}{M. Maaroufi}, \bibinfo{person}{Sang-Hoon Yeo}, \bibinfo{person}{R. Miall}, \bibinfo{person}{Peter Holland}, {and} \bibinfo{person}{Joeseph Galea}.} \bibinfo{year}{2023}\natexlab{}.
\newblock \showarticletitle{A methodological framework to assess the accuracy of virtual reality hand-tracking systems: A case study with the Meta Quest 2}.
\newblock \bibinfo{journal}{\emph{Behavior Research Methods}}  \bibinfo{volume}{56} (\bibinfo{date}{02} \bibinfo{year}{2023}), \bibinfo{pages}{1--12}.
\newblock
\urldef\tempurl%
\url{https://doi.org/10.3758/s13428-022-02051-8}
\showDOI{\tempurl}


\bibitem[Ahuja et~al\mbox{.}(2021)]%
        {Ahuja2021TouchPose}
\bibfield{author}{\bibinfo{person}{Karan Ahuja}, \bibinfo{person}{Paul Streli}, {and} \bibinfo{person}{Christian Holz}.} \bibinfo{year}{2021}\natexlab{}.
\newblock \showarticletitle{TouchPose: Hand Pose Prediction, Depth Estimation, and Touch Classification from Capacitive Images}. In \bibinfo{booktitle}{\emph{The 34th Annual ACM Symposium on User Interface Software and Technology}} (Virtual Event, USA) \emph{(\bibinfo{series}{UIST '21})}. \bibinfo{publisher}{Association for Computing Machinery}, \bibinfo{address}{New York, NY, USA}, \bibinfo{pages}{997–1009}.
\newblock
\showISBNx{9781450386357}
\urldef\tempurl%
\url{https://doi.org/10.1145/3472749.3474801}
\showDOI{\tempurl}


\bibitem[Bhirangi et~al\mbox{.}(2021)]%
        {bhirangi2021reskin}
\bibfield{author}{\bibinfo{person}{Raunaq Bhirangi}, \bibinfo{person}{Tess Hellebrekers}, \bibinfo{person}{Carmel Majidi}, {and} \bibinfo{person}{Abhinav Gupta}.} \bibinfo{year}{2021}\natexlab{}.
\newblock \showarticletitle{ReSkin:versatile, replaceable, lasting tactile skins}. In \bibinfo{booktitle}{\emph{CoRL}}.
\newblock


\bibitem[Bolt(1980)]%
        {Bolt1980VoiceGesture}
\bibfield{author}{\bibinfo{person}{Richard~A. Bolt}.} \bibinfo{year}{1980}\natexlab{}.
\newblock \bibinfo{booktitle}{\emph{“Put-that-there”: Voice and gesture at the graphics interface}}.
\newblock \bibinfo{publisher}{Association for Computing Machinery}, \bibinfo{address}{New York, NY, USA}, \bibinfo{pages}{262–270}.
\newblock
\showISBNx{0897910214}
\urldef\tempurl%
\url{https://doi.org/10.1145/800250.807503}
\showURL{%
\tempurl}


\bibitem[Brahmbhatt(2020)]%
        {Brahmbhatt2020Grasp}
\bibfield{author}{\bibinfo{person}{Samarth Brahmbhatt}.} \bibinfo{year}{2020}\natexlab{}.
\newblock \emph{\bibinfo{title}{Grasp Contact Between Hand and Object: Capture, Analysis, and Applications}}.
\newblock \bibinfo{thesistype}{Ph.\,D. Dissertation}. \bibinfo{school}{Georgia Institute of Technology}, \bibinfo{address}{Atlanta, GA, USA}.
\newblock
\newblock
\shownote{PhD Thesis}.


\bibitem[Brasier et~al\mbox{.}(2020)]%
        {Brasier2020ARPads}
\bibfield{author}{\bibinfo{person}{Eugenie Brasier}, \bibinfo{person}{Olivier Chapuis}, \bibinfo{person}{Nicolas Ferey}, \bibinfo{person}{Jeanne Vezien}, {and} \bibinfo{person}{Caroline Appert}.} \bibinfo{year}{2020}\natexlab{}.
\newblock \showarticletitle{ARPads: Mid-air Indirect Input for Augmented Reality}. In \bibinfo{booktitle}{\emph{2020 IEEE International Symposium on Mixed and Augmented Reality (ISMAR)}}. \bibinfo{pages}{332--343}.
\newblock
\urldef\tempurl%
\url{https://doi.org/10.1109/ISMAR50242.2020.00060}
\showDOI{\tempurl}


\bibitem[Buechley et~al\mbox{.}(2010)]%
        {Buechley2010LivingWall}
\bibfield{author}{\bibinfo{person}{Leah Buechley}, \bibinfo{person}{David Mellis}, \bibinfo{person}{Hannah Perner-Wilson}, \bibinfo{person}{Emily Lovell}, {and} \bibinfo{person}{Bonifaz Kaufmann}.} \bibinfo{year}{2010}\natexlab{}.
\newblock \showarticletitle{Living wall: programmable wallpaper for interactive spaces}. In \bibinfo{booktitle}{\emph{Proceedings of the 18th ACM International Conference on Multimedia}} (Firenze, Italy) \emph{(\bibinfo{series}{MM '10})}. \bibinfo{publisher}{Association for Computing Machinery}, \bibinfo{address}{New York, NY, USA}, \bibinfo{pages}{1401–1402}.
\newblock
\showISBNx{9781605589336}
\urldef\tempurl%
\url{https://doi.org/10.1145/1873951.1874226}
\showDOI{\tempurl}


\bibitem[Cadena et~al\mbox{.}(2016)]%
        {Cadena2016Fingertip}
\bibfield{author}{\bibinfo{person}{Arturo Cadena}, \bibinfo{person}{Rubén Carvajal}, \bibinfo{person}{Bruno Guamán}, \bibinfo{person}{Roger Granda}, \bibinfo{person}{Enrique Peláez}, {and} \bibinfo{person}{Katherine Chiluiza}.} \bibinfo{year}{2016}\natexlab{}.
\newblock \showarticletitle{Fingertip detection approach on depth image sequences for interactive projection system}. In \bibinfo{booktitle}{\emph{2016 IEEE Ecuador Technical Chapters Meeting (ETCM)}}. \bibinfo{pages}{1--6}.
\newblock
\urldef\tempurl%
\url{https://doi.org/10.1109/ETCM.2016.7750827}
\showDOI{\tempurl}


\bibitem[Chen et~al\mbox{.}(2019)]%
        {Chen2019EstimatingFF}
\bibfield{author}{\bibinfo{person}{Nutan Chen}, \bibinfo{person}{G{\"o}ran Westling}, \bibinfo{person}{Benoni~B. Edin}, {and} \bibinfo{person}{Patrick van~der Smagt}.} \bibinfo{year}{2019}\natexlab{}.
\newblock \showarticletitle{Estimating Fingertip Forces, Torques, and Local Curvatures from Fingernail Images}.
\newblock \bibinfo{journal}{\emph{Robotica}}  \bibinfo{volume}{38} (\bibinfo{year}{2019}), \bibinfo{pages}{1242 -- 1262}.
\newblock
\urldef\tempurl%
\url{https://api.semanticscholar.org/CorpusID:202565726}
\showURL{%
\tempurl}


\bibitem[Chen et~al\mbox{.}(2014)]%
        {Chen2014AirTouch}
\bibfield{author}{\bibinfo{person}{Xiang~'Anthony' Chen}, \bibinfo{person}{Julia Schwarz}, \bibinfo{person}{Chris Harrison}, \bibinfo{person}{Jennifer Mankoff}, {and} \bibinfo{person}{Scott~E. Hudson}.} \bibinfo{year}{2014}\natexlab{}.
\newblock \showarticletitle{Air+touch: interweaving touch \& in-air gestures}. In \bibinfo{booktitle}{\emph{Proceedings of the 27th Annual ACM Symposium on User Interface Software and Technology}} (Honolulu, Hawaii, USA) \emph{(\bibinfo{series}{UIST '14})}. \bibinfo{publisher}{Association for Computing Machinery}, \bibinfo{address}{New York, NY, USA}, \bibinfo{pages}{519–525}.
\newblock
\showISBNx{9781450330695}
\urldef\tempurl%
\url{https://doi.org/10.1145/2642918.2647392}
\showDOI{\tempurl}


\bibitem[Cheng et~al\mbox{.}(2022)]%
        {Cheng2022Comfortable}
\bibfield{author}{\bibinfo{person}{Yi~Fei Cheng}, \bibinfo{person}{Tiffany Luong}, \bibinfo{person}{Andreas~Rene Fender}, \bibinfo{person}{Paul Streli}, {and} \bibinfo{person}{Christian Holz}.} \bibinfo{year}{2022}\natexlab{}.
\newblock \showarticletitle{ComforTable User Interfaces: Surfaces Reduce Input Error, Time, and Exertion for Tabletop and Mid-air User Interfaces}. In \bibinfo{booktitle}{\emph{2022 IEEE International Symposium on Mixed and Augmented Reality (ISMAR)}}. \bibinfo{pages}{150--159}.
\newblock
\urldef\tempurl%
\url{https://doi.org/10.1109/ISMAR55827.2022.00029}
\showDOI{\tempurl}


\bibitem[Choi et~al\mbox{.}(2021)]%
        {Choi2021Touchscreens}
\bibfield{author}{\bibinfo{person}{Frederick Choi}, \bibinfo{person}{Sven Mayer}, {and} \bibinfo{person}{Chris Harrison}.} \bibinfo{year}{2021}\natexlab{}.
\newblock \showarticletitle{3D Hand Pose Estimation on Conventional Capacitive Touchscreens}. In \bibinfo{booktitle}{\emph{Proceedings of the 23rd International Conference on Mobile Human-Computer Interaction}} (Toulouse \& Virtual, France) \emph{(\bibinfo{series}{MobileHCI '21})}. \bibinfo{publisher}{Association for Computing Machinery}, \bibinfo{address}{New York, NY, USA}, Article \bibinfo{articleno}{3}, \bibinfo{numpages}{13}~pages.
\newblock
\showISBNx{9781450383288}
\urldef\tempurl%
\url{https://doi.org/10.1145/3447526.3472045}
\showDOI{\tempurl}


\bibitem[Cook and Torrance(1982)]%
        {Cook1982RmodelCG}
\bibfield{author}{\bibinfo{person}{R.~L. Cook} {and} \bibinfo{person}{K.~E. Torrance}.} \bibinfo{year}{1982}\natexlab{}.
\newblock \showarticletitle{A Reflectance Model for Computer Graphics}.
\newblock \bibinfo{journal}{\emph{ACM Trans. Graph.}} \bibinfo{volume}{1}, \bibinfo{number}{1} (\bibinfo{date}{jan} \bibinfo{year}{1982}), \bibinfo{pages}{7–24}.
\newblock
\showISSN{0730-0301}
\urldef\tempurl%
\url{https://doi.org/10.1145/357290.357293}
\showDOI{\tempurl}


\bibitem[Csonka et~al\mbox{.}(2023)]%
        {Csonka2023AIGesture}
\bibfield{author}{\bibinfo{person}{Gergo Csonka}, \bibinfo{person}{Muhammad Khalid}, \bibinfo{person}{Husnain Rafiq}, {and} \bibinfo{person}{Yasir Ali}.} \bibinfo{year}{2023}\natexlab{}.
\newblock \showarticletitle{AI-Based Hand Gesture Recognition Through Camera on Robot}. In \bibinfo{booktitle}{\emph{2023 International Conference on Frontiers of Information Technology (FIT)}}. \bibinfo{pages}{256--261}.
\newblock
\urldef\tempurl%
\url{https://doi.org/10.1109/FIT60620.2023.00054}
\showDOI{\tempurl}


\bibitem[Dobinson et~al\mbox{.}(2022)]%
        {Dobinson2022MicroPress}
\bibfield{author}{\bibinfo{person}{Rhett Dobinson}, \bibinfo{person}{Marc Teyssier}, \bibinfo{person}{J\"{u}rgen Steimle}, {and} \bibinfo{person}{Bruno Fruchard}.} \bibinfo{year}{2022}\natexlab{}.
\newblock \showarticletitle{MicroPress: Detecting Pressure and Hover Distance in Thumb-to-Finger Interactions}. In \bibinfo{booktitle}{\emph{Proceedings of the 2022 ACM Symposium on Spatial User Interaction}} (Online, CA, USA) \emph{(\bibinfo{series}{SUI '22})}. \bibinfo{publisher}{Association for Computing Machinery}, \bibinfo{address}{New York, NY, USA}, Article \bibinfo{articleno}{4}, \bibinfo{numpages}{10}~pages.
\newblock
\showISBNx{9781450399487}
\urldef\tempurl%
\url{https://doi.org/10.1145/3565970.3567698}
\showDOI{\tempurl}


\bibitem[Dupr\'{e} et~al\mbox{.}(2024)]%
        {Dupre2024Tripad}
\bibfield{author}{\bibinfo{person}{Camille Dupr\'{e}}, \bibinfo{person}{Caroline Appert}, \bibinfo{person}{St\'{e}phanie Rey}, \bibinfo{person}{Houssem Saidi}, {and} \bibinfo{person}{Emmanuel Pietriga}.} \bibinfo{year}{2024}\natexlab{}.
\newblock \showarticletitle{TriPad: Touch Input in AR on Ordinary Surfaces with Hand Tracking Only}. In \bibinfo{booktitle}{\emph{Proceedings of the CHI Conference on Human Factors in Computing Systems}} (Honolulu, HI, USA) \emph{(\bibinfo{series}{CHI '24})}. \bibinfo{publisher}{Association for Computing Machinery}, \bibinfo{address}{New York, NY, USA}, Article \bibinfo{articleno}{754}, \bibinfo{numpages}{18}~pages.
\newblock
\showISBNx{9798400703300}
\urldef\tempurl%
\url{https://doi.org/10.1145/3613904.3642323}
\showDOI{\tempurl}


\bibitem[Fan and Xiao(2022)]%
        {Fan2022ReducingLatency}
\bibfield{author}{\bibinfo{person}{Neil~Xu Fan} {and} \bibinfo{person}{Robert Xiao}.} \bibinfo{year}{2022}\natexlab{}.
\newblock \showarticletitle{Reducing the Latency of Touch Tracking on Ad-hoc Surfaces}.
\newblock \bibinfo{journal}{\emph{Proc. ACM Hum.-Comput. Interact.}} \bibinfo{volume}{6}, \bibinfo{number}{ISS}, Article \bibinfo{articleno}{577} (\bibinfo{date}{nov} \bibinfo{year}{2022}), \bibinfo{numpages}{11}~pages.
\newblock
\urldef\tempurl%
\url{https://doi.org/10.1145/3567730}
\showDOI{\tempurl}


\bibitem[Feigin et~al\mbox{.}(2016)]%
        {Feigin2016ReolvingMPI}
\bibfield{author}{\bibinfo{person}{Micha Feigin}, \bibinfo{person}{Ayush Bhandari}, \bibinfo{person}{Shahram Izadi}, \bibinfo{person}{Christoph Rhemann}, \bibinfo{person}{Mirko Schmidt}, {and} \bibinfo{person}{Ramesh Raskar}.} \bibinfo{year}{2016}\natexlab{}.
\newblock \showarticletitle{Resolving Multipath Interference in Kinect: An Inverse Problem Approach}.
\newblock \bibinfo{journal}{\emph{IEEE Sensors Journal}} \bibinfo{volume}{16}, \bibinfo{number}{10} (\bibinfo{year}{2016}), \bibinfo{pages}{3419--3427}.
\newblock
\urldef\tempurl%
\url{https://doi.org/10.1109/JSEN.2015.2421360}
\showDOI{\tempurl}


\bibitem[Fitzpatrick(1975)]%
        {fitzpatrick1975soleil}
\bibfield{author}{\bibinfo{person}{T.~B. Fitzpatrick}.} \bibinfo{year}{1975}\natexlab{}.
\newblock \showarticletitle{Soleil et peau}.
\newblock \bibinfo{journal}{\emph{J Med Esthet}} \bibinfo{volume}{2}, \bibinfo{number}{7} (\bibinfo{year}{1975}), \bibinfo{pages}{33--34}.
\newblock


\bibitem[{Google Research}(2023)]%
        {mediapipe2023handlandmarker}
\bibfield{author}{\bibinfo{person}{{Google Research}}.} \bibinfo{year}{2023}\natexlab{}.
\newblock \bibinfo{title}{MediaPipe Hand Landmarker: Real-time Hand Landmark Detection}.
\newblock \bibinfo{howpublished}{\url{https://ai.google.dev/edge/mediapipe/solutions/vision/hand_landmarker}}.
\newblock
\newblock
\shownote{Accessed: 2024-09-01}.


\bibitem[Grady et~al\mbox{.}(2024)]%
        {Grady2024PressureVision++}
\bibfield{author}{\bibinfo{person}{P. Grady}, \bibinfo{person}{J.~A. Collins}, \bibinfo{person}{C. Tang}, \bibinfo{person}{C.~D. Twigg}, \bibinfo{person}{K. Aneja}, \bibinfo{person}{J. Hays}, {and} \bibinfo{person}{C.~C. Kemp}.} \bibinfo{year}{2024}\natexlab{}.
\newblock \showarticletitle{PressureVision++: Estimating Fingertip Pressure from Diverse RGB Images}. In \bibinfo{booktitle}{\emph{2024 IEEE/CVF Winter Conference on Applications of Computer Vision (WACV)}}. \bibinfo{publisher}{IEEE Computer Society}, \bibinfo{address}{Los Alamitos, CA, USA}, \bibinfo{pages}{8683--8693}.
\newblock
\urldef\tempurl%
\url{https://doi.org/10.1109/WACV57701.2024.00850}
\showDOI{\tempurl}


\bibitem[Grady et~al\mbox{.}(2022)]%
        {Grady2022PressureVision}
\bibfield{author}{\bibinfo{person}{Patrick Grady}, \bibinfo{person}{Chengcheng Tang}, \bibinfo{person}{Samarth Brahmbhatt}, \bibinfo{person}{Christopher~D. Twigg}, \bibinfo{person}{Chengde Wan}, \bibinfo{person}{James Hays}, {and} \bibinfo{person}{Charles~C. Kemp}.} \bibinfo{year}{2022}\natexlab{}.
\newblock \showarticletitle{PressureVision: Estimating Hand Pressure from a Single RGB Image}. In \bibinfo{booktitle}{\emph{Computer Vision – ECCV 2022: 17th European Conference, Tel Aviv, Israel, October 23–27, 2022, Proceedings, Part VI}} (Tel Aviv, Israel). \bibinfo{publisher}{Springer-Verlag}, \bibinfo{address}{Berlin, Heidelberg}, \bibinfo{pages}{328–345}.
\newblock
\showISBNx{978-3-031-20067-0}
\urldef\tempurl%
\url{https://doi.org/10.1007/978-3-031-20068-7_19}
\showDOI{\tempurl}


\bibitem[Gu et~al\mbox{.}(2019)]%
        {Gu2019FingerIMU}
\bibfield{author}{\bibinfo{person}{Yizheng Gu}, \bibinfo{person}{Chun Yu}, \bibinfo{person}{Zhipeng Li}, \bibinfo{person}{Weiqi Li}, \bibinfo{person}{Shuchang Xu}, \bibinfo{person}{Xiaoying Wei}, {and} \bibinfo{person}{Yuanchun Shi}.} \bibinfo{year}{2019}\natexlab{}.
\newblock \showarticletitle{Accurate and Low-Latency Sensing of Touch Contact on Any Surface with Finger-Worn IMU Sensor}. In \bibinfo{booktitle}{\emph{Proceedings of the 32nd Annual ACM Symposium on User Interface Software and Technology}} (New Orleans, LA, USA) \emph{(\bibinfo{series}{UIST '19})}. \bibinfo{publisher}{Association for Computing Machinery}, \bibinfo{address}{New York, NY, USA}, \bibinfo{pages}{1059–1070}.
\newblock
\showISBNx{9781450368162}
\urldef\tempurl%
\url{https://doi.org/10.1145/3332165.3347947}
\showDOI{\tempurl}


\bibitem[Guo et~al\mbox{.}(2015)]%
        {Guo2015CapAuth}
\bibfield{author}{\bibinfo{person}{Anhong Guo}, \bibinfo{person}{Robert Xiao}, {and} \bibinfo{person}{Chris Harrison}.} \bibinfo{year}{2015}\natexlab{}.
\newblock \showarticletitle{CapAuth: Identifying and Differentiating User Handprints on Commodity Capacitive Touchscreens}. In \bibinfo{booktitle}{\emph{Proceedings of the 2015 International Conference on Interactive Tabletops \& Surfaces}} (Madeira, Portugal) \emph{(\bibinfo{series}{ITS '15})}. \bibinfo{publisher}{Association for Computing Machinery}, \bibinfo{address}{New York, NY, USA}, \bibinfo{pages}{59–62}.
\newblock
\showISBNx{9781450338998}
\urldef\tempurl%
\url{https://doi.org/10.1145/2817721.2817722}
\showDOI{\tempurl}


\bibitem[Harrison et~al\mbox{.}(2011)]%
        {Harrison2011Omnitouch}
\bibfield{author}{\bibinfo{person}{Chris Harrison}, \bibinfo{person}{Hrvoje Benko}, {and} \bibinfo{person}{Andrew~D. Wilson}.} \bibinfo{year}{2011}\natexlab{}.
\newblock \showarticletitle{OmniTouch: wearable multitouch interaction everywhere}. In \bibinfo{booktitle}{\emph{Proceedings of the 24th Annual ACM Symposium on User Interface Software and Technology}} (Santa Barbara, California, USA) \emph{(\bibinfo{series}{UIST '11})}. \bibinfo{publisher}{Association for Computing Machinery}, \bibinfo{address}{New York, NY, USA}, \bibinfo{pages}{441–450}.
\newblock
\showISBNx{9781450307161}
\urldef\tempurl%
\url{https://doi.org/10.1145/2047196.2047255}
\showDOI{\tempurl}


\bibitem[Harrison and Hudson(2008)]%
        {Harrison2008Scratch}
\bibfield{author}{\bibinfo{person}{Chris Harrison} {and} \bibinfo{person}{Scott~E. Hudson}.} \bibinfo{year}{2008}\natexlab{}.
\newblock \showarticletitle{Scratch input: creating large, inexpensive, unpowered and mobile finger input surfaces}. In \bibinfo{booktitle}{\emph{Proceedings of the 21st Annual ACM Symposium on User Interface Software and Technology}} (Monterey, CA, USA) \emph{(\bibinfo{series}{UIST '08})}. \bibinfo{publisher}{Association for Computing Machinery}, \bibinfo{address}{New York, NY, USA}, \bibinfo{pages}{205–208}.
\newblock
\showISBNx{9781595939753}
\urldef\tempurl%
\url{https://doi.org/10.1145/1449715.1449747}
\showDOI{\tempurl}


\bibitem[Haubner et~al\mbox{.}(2013)]%
        {Haubner2013Singledepth}
\bibfield{author}{\bibinfo{person}{Nadia Haubner}, \bibinfo{person}{Ulrich Schwanecke}, \bibinfo{person}{Ralf D\"{o}rner}, \bibinfo{person}{Simon Lehmann}, {and} \bibinfo{person}{Johannes Luderschmidt}.} \bibinfo{year}{2013}\natexlab{}.
\newblock \showarticletitle{Detecting interaction above digital tabletops using a single depth camera}.
\newblock \bibinfo{journal}{\emph{Mach. Vision Appl.}} \bibinfo{volume}{24}, \bibinfo{number}{8} (\bibinfo{date}{nov} \bibinfo{year}{2013}), \bibinfo{pages}{1575–1587}.
\newblock
\showISSN{0932-8092}
\urldef\tempurl%
\url{https://doi.org/10.1007/s00138-013-0538-5}
\showDOI{\tempurl}


\bibitem[Hincapi\'{e}-Ramos et~al\mbox{.}(2014)]%
        {Hincapi2014MidAir}
\bibfield{author}{\bibinfo{person}{Juan~David Hincapi\'{e}-Ramos}, \bibinfo{person}{Xiang Guo}, \bibinfo{person}{Paymahn Moghadasian}, {and} \bibinfo{person}{Pourang Irani}.} \bibinfo{year}{2014}\natexlab{}.
\newblock \showarticletitle{Consumed endurance: a metric to quantify arm fatigue of mid-air interactions}. In \bibinfo{booktitle}{\emph{Proceedings of the SIGCHI Conference on Human Factors in Computing Systems}} (Toronto, Ontario, Canada) \emph{(\bibinfo{series}{CHI '14})}. \bibinfo{publisher}{Association for Computing Machinery}, \bibinfo{address}{New York, NY, USA}, \bibinfo{pages}{1063–1072}.
\newblock
\showISBNx{9781450324731}
\urldef\tempurl%
\url{https://doi.org/10.1145/2556288.2557130}
\showDOI{\tempurl}


\bibitem[Hinckley et~al\mbox{.}(2016)]%
        {Hinckley2016PreTouchSF}
\bibfield{author}{\bibinfo{person}{Ken Hinckley}, \bibinfo{person}{Seongkook Heo}, \bibinfo{person}{Michel Pahud}, \bibinfo{person}{Christian Holz}, \bibinfo{person}{Hrvoje Benko}, \bibinfo{person}{Abigail Sellen}, \bibinfo{person}{Richard Banks}, \bibinfo{person}{Kenton~P. O'Hara}, \bibinfo{person}{Gavin Smyth}, {and} \bibinfo{person}{William~A.S. Buxton}.} \bibinfo{year}{2016}\natexlab{}.
\newblock \showarticletitle{Pre-Touch Sensing for Mobile Interaction}.
\newblock \bibinfo{journal}{\emph{Proceedings of the 2016 CHI Conference on Human Factors in Computing Systems}} (\bibinfo{year}{2016}).
\newblock
\urldef\tempurl%
\url{https://api.semanticscholar.org/CorpusID:15482351}
\showURL{%
\tempurl}


\bibitem[Huang et~al\mbox{.}(2023)]%
        {Huang2023VR360Telepresence}
\bibfield{author}{\bibinfo{person}{Xincheng Huang}, \bibinfo{person}{James Riddell}, {and} \bibinfo{person}{Robert Xiao}.} \bibinfo{year}{2023}\natexlab{}.
\newblock \showarticletitle{Virtual Reality Telepresence: 360-Degree Video Streaming with Edge-Compute Assisted Static Foveated Compression}.
\newblock \bibinfo{journal}{\emph{IEEE Transactions on Visualization and Computer Graphics}} \bibinfo{volume}{29}, \bibinfo{number}{11} (\bibinfo{year}{2023}), \bibinfo{pages}{4525--4534}.
\newblock
\urldef\tempurl%
\url{https://doi.org/10.1109/TVCG.2023.3320255}
\showDOI{\tempurl}


\bibitem[Huang and Xiao(2024)]%
        {Huang2024SurfShare}
\bibfield{author}{\bibinfo{person}{Xincheng Huang} {and} \bibinfo{person}{Robert Xiao}.} \bibinfo{year}{2024}\natexlab{}.
\newblock \showarticletitle{SurfShare: Lightweight Spatially Consistent Physical Surface and Virtual Replica Sharing with Head-mounted Mixed-Reality}.
\newblock \bibinfo{journal}{\emph{Proc. ACM Interact. Mob. Wearable Ubiquitous Technol.}} \bibinfo{volume}{7}, \bibinfo{number}{4}, Article \bibinfo{articleno}{162} (\bibinfo{date}{Jan.} \bibinfo{year}{2024}), \bibinfo{numpages}{24}~pages.
\newblock
\urldef\tempurl%
\url{https://doi.org/10.1145/3631418}
\showDOI{\tempurl}


\bibitem[Huang et~al\mbox{.}(2024)]%
        {Huang2024VirtualNexus}
\bibfield{author}{\bibinfo{person}{Xincheng Huang}, \bibinfo{person}{Michael Yin}, \bibinfo{person}{Ziyi Xia}, {and} \bibinfo{person}{Robert Xiao}.} \bibinfo{year}{2024}\natexlab{}.
\newblock \showarticletitle{VirtualNexus: Enhancing 360-Degree Video AR/VR Collaboration with Environment Cutouts and Virtual Replicas}. In \bibinfo{booktitle}{\emph{Proceedings of the 37th Annual ACM Symposium on User Interface Software and Technology}} (Pittsburgh, PA, USA) \emph{(\bibinfo{series}{UIST '24})}. \bibinfo{publisher}{Association for Computing Machinery}, \bibinfo{address}{New York, NY, USA}, Article \bibinfo{articleno}{55}, \bibinfo{numpages}{12}~pages.
\newblock
\showISBNx{9798400706288}
\urldef\tempurl%
\url{https://doi.org/10.1145/3654777.3676377}
\showDOI{\tempurl}


\bibitem[Hwang and Lim(2017)]%
        {Hwang2017InferringIF}
\bibfield{author}{\bibinfo{person}{Wonjun Hwang} {and} \bibinfo{person}{Soo-Chul Lim}.} \bibinfo{year}{2017}\natexlab{}.
\newblock \showarticletitle{Inferring Interaction Force from Visual Information without Using Physical Force Sensors}.
\newblock \bibinfo{journal}{\emph{Sensors (Basel, Switzerland)}}  \bibinfo{volume}{17} (\bibinfo{year}{2017}).
\newblock
\urldef\tempurl%
\url{https://api.semanticscholar.org/CorpusID:12487487}
\showURL{%
\tempurl}


\bibitem[Ishii et~al\mbox{.}(1999)]%
        {Ishii1999PingPongPlus}
\bibfield{author}{\bibinfo{person}{Hiroshi Ishii}, \bibinfo{person}{Craig Wisneski}, \bibinfo{person}{Julian Orbanes}, \bibinfo{person}{Ben Chun}, {and} \bibinfo{person}{Joe Paradiso}.} \bibinfo{year}{1999}\natexlab{}.
\newblock \showarticletitle{PingPongPlus: design of an athletic-tangible interface for computer-supported cooperative play}. In \bibinfo{booktitle}{\emph{Proceedings of the SIGCHI Conference on Human Factors in Computing Systems}} (Pittsburgh, Pennsylvania, USA) \emph{(\bibinfo{series}{CHI '99})}. \bibinfo{publisher}{Association for Computing Machinery}, \bibinfo{address}{New York, NY, USA}, \bibinfo{pages}{394–401}.
\newblock
\showISBNx{0201485591}
\urldef\tempurl%
\url{https://doi.org/10.1145/302979.303115}
\showDOI{\tempurl}


\bibitem[Kamel(1990)]%
        {Kamel1990}
\bibfield{author}{\bibinfo{person}{Ragui Kamel}.} \bibinfo{year}{1990}\natexlab{}.
\newblock \showarticletitle{Guest Editor's Introduction: Voice in Computing}.
\newblock \bibinfo{journal}{\emph{Computer}} \bibinfo{volume}{23}, \bibinfo{number}{8} (\bibinfo{date}{aug} \bibinfo{year}{1990}), \bibinfo{pages}{8–9}.
\newblock
\showISSN{0018-9162}


\bibitem[Kim et~al\mbox{.}(2011)]%
        {Kim2011CapacitiveTS}
\bibfield{author}{\bibinfo{person}{Hongki Kim}, \bibinfo{person}{Seung-Gun Lee}, {and} \bibinfo{person}{Kwang-Seok Yun}.} \bibinfo{year}{2011}\natexlab{}.
\newblock \showarticletitle{Capacitive tactile sensor array for touch screen application}.
\newblock \bibinfo{journal}{\emph{Sensors and Actuators A-physical}}  \bibinfo{volume}{165} (\bibinfo{year}{2011}), \bibinfo{pages}{2--7}.
\newblock
\urldef\tempurl%
\url{https://api.semanticscholar.org/CorpusID:109271148}
\showURL{%
\tempurl}


\bibitem[Kurz(2014)]%
        {Kurz2014ThermalTouch}
\bibfield{author}{\bibinfo{person}{Daniel Kurz}.} \bibinfo{year}{2014}\natexlab{}.
\newblock \showarticletitle{Thermal touch: Thermography-enabled everywhere touch interfaces for mobile augmented reality applications}. In \bibinfo{booktitle}{\emph{2014 IEEE International Symposium on Mixed and Augmented Reality (ISMAR)}}. \bibinfo{pages}{9--16}.
\newblock
\urldef\tempurl%
\url{https://doi.org/10.1109/ISMAR.2014.6948403}
\showDOI{\tempurl}


\bibitem[Lange and Seitz(2001)]%
        {Lange2001TOFCamera}
\bibfield{author}{\bibinfo{person}{R. Lange} {and} \bibinfo{person}{P. Seitz}.} \bibinfo{year}{2001}\natexlab{}.
\newblock \showarticletitle{Solid-state time-of-flight range camera}.
\newblock \bibinfo{journal}{\emph{IEEE Journal of Quantum Electronics}} \bibinfo{volume}{37}, \bibinfo{number}{3} (\bibinfo{year}{2001}), \bibinfo{pages}{390--397}.
\newblock
\urldef\tempurl%
\url{https://doi.org/10.1109/3.910448}
\showDOI{\tempurl}


\bibitem[Liang et~al\mbox{.}(2023)]%
        {Liang2023ShadowTouch}
\bibfield{author}{\bibinfo{person}{Chen Liang}, \bibinfo{person}{Xutong Wang}, \bibinfo{person}{Zisu Li}, \bibinfo{person}{Chi Hsia}, \bibinfo{person}{Mingming Fan}, \bibinfo{person}{Chun Yu}, {and} \bibinfo{person}{Yuanchun Shi}.} \bibinfo{year}{2023}\natexlab{}.
\newblock \showarticletitle{ShadowTouch: Enabling Free-Form Touch-Based Hand-to-Surface Interaction with Wrist-Mounted Illuminant by Shadow Projection}. In \bibinfo{booktitle}{\emph{Proceedings of the 36th Annual ACM Symposium on User Interface Software and Technology}} (San Francisco, CA, USA) \emph{(\bibinfo{series}{UIST '23})}. \bibinfo{publisher}{Association for Computing Machinery}, \bibinfo{address}{New York, NY, USA}, Article \bibinfo{articleno}{27}, \bibinfo{numpages}{14}~pages.
\newblock
\showISBNx{9798400701320}
\urldef\tempurl%
\url{https://doi.org/10.1145/3586183.3606785}
\showDOI{\tempurl}


\bibitem[Lindeman et~al\mbox{.}(1999)]%
        {Lindeman1999TowardsVR}
\bibfield{author}{\bibinfo{person}{Robert~W. Lindeman}, \bibinfo{person}{John~L. Sibert}, {and} \bibinfo{person}{James~K. Hahn}.} \bibinfo{year}{1999}\natexlab{}.
\newblock \showarticletitle{Towards usable VR: an empirical study of user interfaces for immersive virtual environments}. In \bibinfo{booktitle}{\emph{Proceedings of the SIGCHI Conference on Human Factors in Computing Systems}} (Pittsburgh, Pennsylvania, USA) \emph{(\bibinfo{series}{CHI '99})}. \bibinfo{publisher}{Association for Computing Machinery}, \bibinfo{address}{New York, NY, USA}, \bibinfo{pages}{64–71}.
\newblock
\showISBNx{0201485591}
\urldef\tempurl%
\url{https://doi.org/10.1145/302979.302995}
\showDOI{\tempurl}


\bibitem[Ma et~al\mbox{.}(2015)]%
        {Ma2015HapticKF}
\bibfield{author}{\bibinfo{person}{Zhaoyuan Ma}, \bibinfo{person}{Darren Edge}, \bibinfo{person}{Leah Findlater}, {and} \bibinfo{person}{Hong~Z. Tan}.} \bibinfo{year}{2015}\natexlab{}.
\newblock \showarticletitle{Haptic keyclick feedback improves typing speed and reduces typing errors on a flat keyboard}.
\newblock \bibinfo{journal}{\emph{2015 IEEE World Haptics Conference (WHC)}} (\bibinfo{year}{2015}), \bibinfo{pages}{220--227}.
\newblock
\urldef\tempurl%
\url{https://api.semanticscholar.org/CorpusID:15523816}
\showURL{%
\tempurl}


\bibitem[MacKenzie and Soukoreff(2003)]%
        {MackKenzie2003PhraseSet}
\bibfield{author}{\bibinfo{person}{I.~Scott MacKenzie} {and} \bibinfo{person}{R.~William Soukoreff}.} \bibinfo{year}{2003}\natexlab{}.
\newblock \showarticletitle{Phrase sets for evaluating text entry techniques}. In \bibinfo{booktitle}{\emph{CHI '03 Extended Abstracts on Human Factors in Computing Systems}} (Ft. Lauderdale, Florida, USA) \emph{(\bibinfo{series}{CHI EA '03})}. \bibinfo{publisher}{Association for Computing Machinery}, \bibinfo{address}{New York, NY, USA}, \bibinfo{pages}{754–755}.
\newblock
\showISBNx{1581136374}
\urldef\tempurl%
\url{https://doi.org/10.1145/765891.765971}
\showDOI{\tempurl}


\bibitem[Mascaro and Asada(2004)]%
        {Mascaro2004MeasurementOF}
\bibfield{author}{\bibinfo{person}{Stephen~A. Mascaro} {and} \bibinfo{person}{Haruhiko Asada}.} \bibinfo{year}{2004}\natexlab{}.
\newblock \showarticletitle{Measurement of finger posture and three-axis fingertip touch force using fingernail sensors}.
\newblock \bibinfo{journal}{\emph{IEEE Transactions on Robotics and Automation}}  \bibinfo{volume}{20} (\bibinfo{year}{2004}), \bibinfo{pages}{26--35}.
\newblock
\urldef\tempurl%
\url{https://api.semanticscholar.org/CorpusID:7842621}
\showURL{%
\tempurl}


\bibitem[Mascaro and Asada(2001)]%
        {Mascaro2001PhotoplethysmographFS}
\bibfield{author}{\bibinfo{person}{Stephen~A. Mascaro} {and} \bibinfo{person}{H.~Harry Asada}.} \bibinfo{year}{2001}\natexlab{}.
\newblock \showarticletitle{Photoplethysmograph fingernail sensors for measuring finger forces without haptic obstruction}.
\newblock \bibinfo{journal}{\emph{IEEE Trans. Robotics Autom.}}  \bibinfo{volume}{17} (\bibinfo{year}{2001}), \bibinfo{pages}{698--708}.
\newblock
\urldef\tempurl%
\url{https://api.semanticscholar.org/CorpusID:17720714}
\showURL{%
\tempurl}


\bibitem[Matulic et~al\mbox{.}(2023)]%
        {Matulic2023AboveScreen}
\bibfield{author}{\bibinfo{person}{Fabrice Matulic}, \bibinfo{person}{Taiga Kashima}, \bibinfo{person}{Deniz Beker}, \bibinfo{person}{Daichi Suzuo}, \bibinfo{person}{Hiroshi Fujiwara}, {and} \bibinfo{person}{Daniel Vogel}.} \bibinfo{year}{2023}\natexlab{}.
\newblock \showarticletitle{Above-Screen Fingertip Tracking with a Phone in Virtual Reality}. In \bibinfo{booktitle}{\emph{Extended Abstracts of the 2023 CHI Conference on Human Factors in Computing Systems}} (Hamburg, Germany) \emph{(\bibinfo{series}{CHI EA '23})}. \bibinfo{publisher}{Association for Computing Machinery}, \bibinfo{address}{New York, NY, USA}, Article \bibinfo{articleno}{18}, \bibinfo{numpages}{7}~pages.
\newblock
\showISBNx{9781450394222}
\urldef\tempurl%
\url{https://doi.org/10.1145/3544549.3585728}
\showDOI{\tempurl}


\bibitem[Medeiros et~al\mbox{.}(2023)]%
        {Medeiros2023Beniefits}
\bibfield{author}{\bibinfo{person}{Daniel Medeiros}, \bibinfo{person}{Graham Wilson}, \bibinfo{person}{Mark Mcgill}, {and} \bibinfo{person}{Stephen~Anthony Brewster}.} \bibinfo{year}{2023}\natexlab{}.
\newblock \showarticletitle{The Benefits of Passive Haptics and Perceptual Manipulation for Extended Reality Interactions in Constrained Passenger Spaces}. In \bibinfo{booktitle}{\emph{Proceedings of the 2023 CHI Conference on Human Factors in Computing Systems}} (Hamburg, Germany) \emph{(\bibinfo{series}{CHI '23})}. \bibinfo{publisher}{Association for Computing Machinery}, \bibinfo{address}{New York, NY, USA}, Article \bibinfo{articleno}{232}, \bibinfo{numpages}{19}~pages.
\newblock
\showISBNx{9781450394215}
\urldef\tempurl%
\url{https://doi.org/10.1145/3544548.3581079}
\showDOI{\tempurl}


\bibitem[Meier et~al\mbox{.}(2021)]%
        {Meier2021TapID}
\bibfield{author}{\bibinfo{person}{M. Meier}, \bibinfo{person}{P. Streli}, \bibinfo{person}{A. Fender}, {and} \bibinfo{person}{C. Holz}.} \bibinfo{year}{2021}\natexlab{}.
\newblock \showarticletitle{TapID: Rapid Touch Interaction in Virtual Reality using Wearable Sensing}. In \bibinfo{booktitle}{\emph{2021 IEEE Virtual Reality and 3D User Interfaces (VR)}}. \bibinfo{publisher}{IEEE Computer Society}, \bibinfo{address}{Los Alamitos, CA, USA}, \bibinfo{pages}{519--528}.
\newblock
\urldef\tempurl%
\url{https://doi.org/10.1109/VR50410.2021.00076}
\showDOI{\tempurl}


\bibitem[{Microsoft Corporation}(2019)]%
        {azurekinectdk2019}
\bibfield{author}{\bibinfo{person}{{Microsoft Corporation}}.} \bibinfo{year}{2019}\natexlab{}.
\newblock \bibinfo{title}{{Azure Kinect DK}}.
\newblock \bibinfo{howpublished}{\url{https://learn.microsoft.com/en-us/previous-versions/azure/kinect-dk/}}.
\newblock
\newblock
\shownote{Accessed: 2024-09-01}.


\bibitem[Mollyn and Harrison(2024)]%
        {Mollyn2024EgoTouch}
\bibfield{author}{\bibinfo{person}{Vimal Mollyn} {and} \bibinfo{person}{Chris Harrison}.} \bibinfo{year}{2024}\natexlab{}.
\newblock \showarticletitle{EgoTouch: On-Body Touch Input Using AR/VR Headset Cameras}. In \bibinfo{booktitle}{\emph{Proceedings of the 37th Annual ACM Symposium on User Interface Software and Technology}} (Pittsburgh, PA, USA) \emph{(\bibinfo{series}{UIST '24})}. \bibinfo{publisher}{Association for Computing Machinery}, \bibinfo{address}{New York, NY, USA}, Article \bibinfo{articleno}{69}, \bibinfo{numpages}{11}~pages.
\newblock
\showISBNx{9798400706288}
\urldef\tempurl%
\url{https://doi.org/10.1145/3654777.3676455}
\showDOI{\tempurl}


\bibitem[Oh et~al\mbox{.}(2020)]%
        {Oh2020FingerTouch}
\bibfield{author}{\bibinfo{person}{Ju~Young Oh}, \bibinfo{person}{Ji-Hyung Park}, {and} \bibinfo{person}{Jung-Min Park}.} \bibinfo{year}{2020}\natexlab{}.
\newblock \showarticletitle{FingerTouch: Touch Interaction Using a Fingernail-Mounted Sensor on a Head-Mounted Display for Augmented Reality}.
\newblock \bibinfo{journal}{\emph{IEEE Access}}  \bibinfo{volume}{8} (\bibinfo{year}{2020}), \bibinfo{pages}{101192--101208}.
\newblock
\urldef\tempurl%
\url{https://doi.org/10.1109/ACCESS.2020.2997972}
\showDOI{\tempurl}


\bibitem[Paradiso et~al\mbox{.}(2002)]%
        {Paradiso2002AccousticKnocks}
\bibfield{author}{\bibinfo{person}{J.A. Paradiso}, \bibinfo{person}{Che~King Leo}, \bibinfo{person}{N. Checka}, {and} \bibinfo{person}{Kaijen Hsiao}.} \bibinfo{year}{2002}\natexlab{}.
\newblock \showarticletitle{Passive acoustic sensing for tracking knocks atop large interactive displays}. In \bibinfo{booktitle}{\emph{SENSORS, 2002 IEEE}}, Vol.~\bibinfo{volume}{1}. \bibinfo{pages}{521--527 vol.1}.
\newblock
\urldef\tempurl%
\url{https://doi.org/10.1109/ICSENS.2002.1037150}
\showDOI{\tempurl}


\bibitem[Pei et~al\mbox{.}(2022)]%
        {Pei2022ForceSight}
\bibfield{author}{\bibinfo{person}{Siyou Pei}, \bibinfo{person}{Pradyumna Chari}, \bibinfo{person}{Xue Wang}, \bibinfo{person}{Xiaoying Yang}, \bibinfo{person}{Achuta Kadambi}, {and} \bibinfo{person}{Yang Zhang}.} \bibinfo{year}{2022}\natexlab{}.
\newblock \showarticletitle{ForceSight: Non-Contact Force Sensing with Laser Speckle Imaging}. In \bibinfo{booktitle}{\emph{Proceedings of the 35th Annual ACM Symposium on User Interface Software and Technology}} (Bend, OR, USA) \emph{(\bibinfo{series}{UIST '22})}. \bibinfo{publisher}{Association for Computing Machinery}, \bibinfo{address}{New York, NY, USA}, Article \bibinfo{articleno}{25}, \bibinfo{numpages}{11}~pages.
\newblock
\showISBNx{9781450393201}
\urldef\tempurl%
\url{https://doi.org/10.1145/3526113.3545622}
\showDOI{\tempurl}


\bibitem[Pham et~al\mbox{.}(2018)]%
        {Pham2018HandObjectCF}
\bibfield{author}{\bibinfo{person}{Tu-Hoa Pham}, \bibinfo{person}{Nikolaos Kyriazis}, \bibinfo{person}{Antonis~A. Argyros}, {and} \bibinfo{person}{Abderrahmane Kheddar}.} \bibinfo{year}{2018}\natexlab{}.
\newblock \showarticletitle{Hand-Object Contact Force Estimation from Markerless Visual Tracking}.
\newblock \bibinfo{journal}{\emph{IEEE Transactions on Pattern Analysis and Machine Intelligence}}  \bibinfo{volume}{40} (\bibinfo{year}{2018}), \bibinfo{pages}{2883--2896}.
\newblock
\urldef\tempurl%
\url{https://api.semanticscholar.org/CorpusID:51612933}
\showURL{%
\tempurl}


\bibitem[Rendl et~al\mbox{.}(2014)]%
        {Rendl2014Presstures}
\bibfield{author}{\bibinfo{person}{Christian Rendl}, \bibinfo{person}{Patrick Greindl}, \bibinfo{person}{Kathrin Probst}, \bibinfo{person}{Martin Behrens}, {and} \bibinfo{person}{Michael Haller}.} \bibinfo{year}{2014}\natexlab{}.
\newblock \showarticletitle{Presstures: exploring pressure-sensitive multi-touch gestures on trackpads}. In \bibinfo{booktitle}{\emph{Proceedings of the SIGCHI Conference on Human Factors in Computing Systems}} (Toronto, Ontario, Canada) \emph{(\bibinfo{series}{CHI '14})}. \bibinfo{publisher}{Association for Computing Machinery}, \bibinfo{address}{New York, NY, USA}, \bibinfo{pages}{431–434}.
\newblock
\showISBNx{9781450324731}
\urldef\tempurl%
\url{https://doi.org/10.1145/2556288.2557146}
\showDOI{\tempurl}


\bibitem[Schmandt and Hulteen(1982)]%
        {Schmandt1982VoiceInterface}
\bibfield{author}{\bibinfo{person}{Christopher Schmandt} {and} \bibinfo{person}{Eric~A. Hulteen}.} \bibinfo{year}{1982}\natexlab{}.
\newblock \showarticletitle{The intelligent voice-interactive interface}. In \bibinfo{booktitle}{\emph{Proceedings of the 1982 Conference on Human Factors in Computing Systems}} (Gaithersburg, Maryland, USA) \emph{(\bibinfo{series}{CHI '82})}. \bibinfo{publisher}{Association for Computing Machinery}, \bibinfo{address}{New York, NY, USA}, \bibinfo{pages}{363–366}.
\newblock
\showISBNx{9781450373890}
\urldef\tempurl%
\url{https://doi.org/10.1145/800049.801812}
\showDOI{\tempurl}


\bibitem[Shen et~al\mbox{.}(2021)]%
        {Shen2021Farout}
\bibfield{author}{\bibinfo{person}{Vivian Shen}, \bibinfo{person}{James Spann}, {and} \bibinfo{person}{Chris Harrison}.} \bibinfo{year}{2021}\natexlab{}.
\newblock \showarticletitle{FarOut Touch: Extending the Range of ad hoc Touch Sensing with Depth Cameras}. In \bibinfo{booktitle}{\emph{Proceedings of the 2021 ACM Symposium on Spatial User Interaction}} (Virtual Event, USA) \emph{(\bibinfo{series}{SUI '21})}. \bibinfo{publisher}{Association for Computing Machinery}, \bibinfo{address}{New York, NY, USA}, Article \bibinfo{articleno}{5}, \bibinfo{numpages}{12}~pages.
\newblock
\showISBNx{9781450390910}
\urldef\tempurl%
\url{https://doi.org/10.1145/3485279.3485281}
\showDOI{\tempurl}


\bibitem[Shi et~al\mbox{.}(2020)]%
        {Shi2020FingerIMU}
\bibfield{author}{\bibinfo{person}{Yilei Shi}, \bibinfo{person}{Haimo Zhang}, \bibinfo{person}{Kaixing Zhao}, \bibinfo{person}{Jiashuo Cao}, \bibinfo{person}{Mengmeng Sun}, {and} \bibinfo{person}{Suranga Nanayakkara}.} \bibinfo{year}{2020}\natexlab{}.
\newblock \showarticletitle{Ready, Steady, Touch! Sensing Physical Contact with a Finger-Mounted IMU}.
\newblock \bibinfo{journal}{\emph{Proc. ACM Interact. Mob. Wearable Ubiquitous Technol.}} \bibinfo{volume}{4}, \bibinfo{number}{2}, Article \bibinfo{articleno}{59} (\bibinfo{date}{jun} \bibinfo{year}{2020}), \bibinfo{numpages}{25}~pages.
\newblock
\urldef\tempurl%
\url{https://doi.org/10.1145/3397309}
\showDOI{\tempurl}


\bibitem[Streli et~al\mbox{.}(2023)]%
        {Streli2023LightSpeckle}
\bibfield{author}{\bibinfo{person}{Paul Streli}, \bibinfo{person}{Jiaxi Jiang}, \bibinfo{person}{Juliete Rossie}, {and} \bibinfo{person}{Christian Holz}.} \bibinfo{year}{2023}\natexlab{}.
\newblock \showarticletitle{Structured Light Speckle: Joint Ego-Centric Depth Estimation and Low-Latency Contact Detection via Remote Vibrometry}. In \bibinfo{booktitle}{\emph{Proceedings of the 36th Annual ACM Symposium on User Interface Software and Technology}} (San Francisco, CA, USA) \emph{(\bibinfo{series}{UIST '23})}. \bibinfo{publisher}{Association for Computing Machinery}, \bibinfo{address}{New York, NY, USA}, Article \bibinfo{articleno}{26}, \bibinfo{numpages}{12}~pages.
\newblock
\showISBNx{9798400701320}
\urldef\tempurl%
\url{https://doi.org/10.1145/3586183.3606749}
\showDOI{\tempurl}


\bibitem[Streli et~al\mbox{.}(2024)]%
        {Streli2024TouchInsight}
\bibfield{author}{\bibinfo{person}{Paul Streli}, \bibinfo{person}{Mark Richardson}, \bibinfo{person}{Fadi Botros}, \bibinfo{person}{Shugao Ma}, \bibinfo{person}{Robert Wang}, {and} \bibinfo{person}{Christian Holz}.} \bibinfo{year}{2024}\natexlab{}.
\newblock \showarticletitle{TouchInsight: Uncertainty-aware Rapid Touch and Text Input for Mixed Reality from Egocentric Vision}. In \bibinfo{booktitle}{\emph{Proceedings of the 37th Annual ACM Symposium on User Interface Software and Technology}} (Pittsburgh, PA, USA) \emph{(\bibinfo{series}{UIST '24})}. \bibinfo{publisher}{Association for Computing Machinery}, \bibinfo{address}{New York, NY, USA}, Article \bibinfo{articleno}{7}, \bibinfo{numpages}{16}~pages.
\newblock
\showISBNx{9798400706288}
\urldef\tempurl%
\url{https://doi.org/10.1145/3654777.3676330}
\showDOI{\tempurl}


\bibitem[Vatavu(2023)]%
        {Vatavu2023Gestures}
\bibfield{author}{\bibinfo{person}{Radu-Daniel Vatavu}.} \bibinfo{year}{2023}\natexlab{}.
\newblock \showarticletitle{iFAD Gestures: Understanding Users’ Gesture Input Performance with Index-Finger Augmentation Devices}. In \bibinfo{booktitle}{\emph{Proceedings of the 2023 CHI Conference on Human Factors in Computing Systems}} (Hamburg, Germany) \emph{(\bibinfo{series}{CHI '23})}. \bibinfo{publisher}{Association for Computing Machinery}, \bibinfo{address}{New York, NY, USA}, Article \bibinfo{articleno}{576}, \bibinfo{numpages}{17}~pages.
\newblock
\showISBNx{9781450394215}
\urldef\tempurl%
\url{https://doi.org/10.1145/3544548.3580928}
\showDOI{\tempurl}


\bibitem[Wilson(2010)]%
        {Wilson2010UsingAD}
\bibfield{author}{\bibinfo{person}{Andrew~D. Wilson}.} \bibinfo{year}{2010}\natexlab{}.
\newblock \showarticletitle{Using a depth camera as a touch sensor}. In \bibinfo{booktitle}{\emph{International Conference on Intelligent Tutoring Systems}}.
\newblock
\urldef\tempurl%
\url{https://api.semanticscholar.org/CorpusID:1333692}
\showURL{%
\tempurl}


\bibitem[Xiao(2018)]%
        {Xiao2018OnWorld}
\bibfield{author}{\bibinfo{person}{Bo Xiao}.} \bibinfo{year}{2018}\natexlab{}.
\newblock \showarticletitle{{On-World Computing: Enabling Interaction on Everyday Surfaces}}.
\newblock  (\bibinfo{date}{10} \bibinfo{year}{2018}).
\newblock
\urldef\tempurl%
\url{https://doi.org/10.1184/R1/7195202.v1}
\showDOI{\tempurl}


\bibitem[Xiao et~al\mbox{.}(2013)]%
        {WorldKit2013Xiao}
\bibfield{author}{\bibinfo{person}{Robert Xiao}, \bibinfo{person}{Chris Harrison}, {and} \bibinfo{person}{Scott~E. Hudson}.} \bibinfo{year}{2013}\natexlab{}.
\newblock \showarticletitle{WorldKit: rapid and easy creation of ad-hoc interactive applications on everyday surfaces}. In \bibinfo{booktitle}{\emph{Proceedings of the SIGCHI Conference on Human Factors in Computing Systems}} (Paris, France) \emph{(\bibinfo{series}{CHI '13})}. \bibinfo{publisher}{Association for Computing Machinery}, \bibinfo{address}{New York, NY, USA}, \bibinfo{pages}{879–888}.
\newblock
\showISBNx{9781450318990}
\urldef\tempurl%
\url{https://doi.org/10.1145/2470654.2466113}
\showDOI{\tempurl}


\bibitem[Xiao et~al\mbox{.}(2016)]%
        {Xiao2016Direct}
\bibfield{author}{\bibinfo{person}{Robert Xiao}, \bibinfo{person}{Scott Hudson}, {and} \bibinfo{person}{Chris Harrison}.} \bibinfo{year}{2016}\natexlab{}.
\newblock \showarticletitle{DIRECT: Making Touch Tracking on Ordinary Surfaces Practical with Hybrid Depth-Infrared Sensing}. In \bibinfo{booktitle}{\emph{Proceedings of the 2016 ACM International Conference on Interactive Surfaces and Spaces}} (Niagara Falls, Ontario, Canada) \emph{(\bibinfo{series}{ISS '16})}. \bibinfo{publisher}{Association for Computing Machinery}, \bibinfo{address}{New York, NY, USA}, \bibinfo{pages}{85–94}.
\newblock
\showISBNx{9781450342483}
\urldef\tempurl%
\url{https://doi.org/10.1145/2992154.2992173}
\showDOI{\tempurl}


\bibitem[Xiao et~al\mbox{.}(2014)]%
        {Xiao2014Toffee}
\bibfield{author}{\bibinfo{person}{Robert Xiao}, \bibinfo{person}{Greg Lew}, \bibinfo{person}{James Marsanico}, \bibinfo{person}{Divya Hariharan}, \bibinfo{person}{Scott Hudson}, {and} \bibinfo{person}{Chris Harrison}.} \bibinfo{year}{2014}\natexlab{}.
\newblock \showarticletitle{Toffee: enabling ad hoc, around-device interaction with acoustic time-of-arrival correlation}. In \bibinfo{booktitle}{\emph{Proceedings of the 16th International Conference on Human-Computer Interaction with Mobile Devices \& Services}} (Toronto, ON, Canada) \emph{(\bibinfo{series}{MobileHCI '14})}. \bibinfo{publisher}{Association for Computing Machinery}, \bibinfo{address}{New York, NY, USA}, \bibinfo{pages}{67–76}.
\newblock
\showISBNx{9781450330046}
\urldef\tempurl%
\url{https://doi.org/10.1145/2628363.2628383}
\showDOI{\tempurl}


\bibitem[Xiao et~al\mbox{.}(2018)]%
        {Xiao2018MRTouch}
\bibfield{author}{\bibinfo{person}{Robert Xiao}, \bibinfo{person}{Julia Schwarz}, \bibinfo{person}{Nick Throm}, \bibinfo{person}{Andrew~D. Wilson}, {and} \bibinfo{person}{Hrvoje Benko}.} \bibinfo{year}{2018}\natexlab{}.
\newblock \showarticletitle{MRTouch: Adding Touch Input to Head-Mounted Mixed Reality}.
\newblock \bibinfo{journal}{\emph{IEEE Transactions on Visualization and Computer Graphics}} \bibinfo{volume}{24}, \bibinfo{number}{4} (\bibinfo{year}{2018}), \bibinfo{pages}{1653--1660}.
\newblock
\urldef\tempurl%
\url{https://doi.org/10.1109/TVCG.2018.2794222}
\showDOI{\tempurl}


\bibitem[Yeo et~al\mbox{.}(2023)]%
        {Yeo2023OmniSense}
\bibfield{author}{\bibinfo{person}{Hui-Shyong Yeo}, \bibinfo{person}{Erwin Wu}, \bibinfo{person}{Daehwa Kim}, \bibinfo{person}{Juyoung Lee}, \bibinfo{person}{Hyung-il Kim}, \bibinfo{person}{Seo~Young Oh}, \bibinfo{person}{Luna Takagi}, \bibinfo{person}{Woontack Woo}, \bibinfo{person}{Hideki Koike}, {and} \bibinfo{person}{Aaron~John Quigley}.} \bibinfo{year}{2023}\natexlab{}.
\newblock \showarticletitle{OmniSense: Exploring Novel Input Sensing and Interaction Techniques on Mobile Device with an Omni-Directional Camera}. In \bibinfo{booktitle}{\emph{Proceedings of the 2023 CHI Conference on Human Factors in Computing Systems}} (Hamburg, Germany) \emph{(\bibinfo{series}{CHI '23})}. \bibinfo{publisher}{Association for Computing Machinery}, \bibinfo{address}{New York, NY, USA}, Article \bibinfo{articleno}{530}, \bibinfo{numpages}{18}~pages.
\newblock
\showISBNx{9781450394215}
\urldef\tempurl%
\url{https://doi.org/10.1145/3544548.3580747}
\showDOI{\tempurl}


\bibitem[Zhang and Wobbrock(2019)]%
        {Zhang2019ByondInput}
\bibfield{author}{\bibinfo{person}{Mingrui~Ray Zhang} {and} \bibinfo{person}{Jacob~O. Wobbrock}.} \bibinfo{year}{2019}\natexlab{}.
\newblock \showarticletitle{Beyond the Input Stream: Making Text Entry Evaluations More Flexible with Transcription Sequences}. In \bibinfo{booktitle}{\emph{Proceedings of the 32nd Annual ACM Symposium on User Interface Software and Technology}} (New Orleans, LA, USA) \emph{(\bibinfo{series}{UIST '19})}. \bibinfo{publisher}{Association for Computing Machinery}, \bibinfo{address}{New York, NY, USA}, \bibinfo{pages}{831–842}.
\newblock
\showISBNx{9781450368162}
\urldef\tempurl%
\url{https://doi.org/10.1145/3332165.3347922}
\showDOI{\tempurl}


\bibitem[Zhang et~al\mbox{.}(2017)]%
        {Zhang2017Electrick}
\bibfield{author}{\bibinfo{person}{Yang Zhang}, \bibinfo{person}{Gierad Laput}, {and} \bibinfo{person}{Chris Harrison}.} \bibinfo{year}{2017}\natexlab{}.
\newblock \showarticletitle{Electrick: Low-Cost Touch Sensing Using Electric Field Tomography}. In \bibinfo{booktitle}{\emph{Proceedings of the 2017 CHI Conference on Human Factors in Computing Systems}} (Denver, Colorado, USA) \emph{(\bibinfo{series}{CHI '17})}. \bibinfo{publisher}{Association for Computing Machinery}, \bibinfo{address}{New York, NY, USA}, \bibinfo{pages}{1–14}.
\newblock
\showISBNx{9781450346559}
\urldef\tempurl%
\url{https://doi.org/10.1145/3025453.3025842}
\showDOI{\tempurl}


\bibitem[Zhang et~al\mbox{.}(2018)]%
        {Zhang2018Wall++}
\bibfield{author}{\bibinfo{person}{Yang Zhang}, \bibinfo{person}{Chouchang~(Jack) Yang}, \bibinfo{person}{Scott~E. Hudson}, \bibinfo{person}{Chris Harrison}, {and} \bibinfo{person}{Alanson Sample}.} \bibinfo{year}{2018}\natexlab{}.
\newblock \showarticletitle{Wall++: Room-Scale Interactive and Context-Aware Sensing}. In \bibinfo{booktitle}{\emph{Proceedings of the 2018 CHI Conference on Human Factors in Computing Systems}} (Montreal QC, Canada) \emph{(\bibinfo{series}{CHI '18})}. \bibinfo{publisher}{Association for Computing Machinery}, \bibinfo{address}{New York, NY, USA}, \bibinfo{pages}{1–15}.
\newblock
\showISBNx{9781450356206}
\urldef\tempurl%
\url{https://doi.org/10.1145/3173574.3173847}
\showDOI{\tempurl}


\bibitem[Zhang et~al\mbox{.}(2016)]%
        {Zhang2016SkinTrack}
\bibfield{author}{\bibinfo{person}{Yang Zhang}, \bibinfo{person}{Junhan Zhou}, \bibinfo{person}{Gierad Laput}, {and} \bibinfo{person}{Chris Harrison}.} \bibinfo{year}{2016}\natexlab{}.
\newblock \showarticletitle{SkinTrack: Using the Body as an Electrical Waveguide for Continuous Finger Tracking on the Skin}. In \bibinfo{booktitle}{\emph{Proceedings of the 2016 CHI Conference on Human Factors in Computing Systems}} (San Jose, California, USA) \emph{(\bibinfo{series}{CHI '16})}. \bibinfo{publisher}{Association for Computing Machinery}, \bibinfo{address}{New York, NY, USA}, \bibinfo{pages}{1491–1503}.
\newblock
\showISBNx{9781450333627}
\urldef\tempurl%
\url{https://doi.org/10.1145/2858036.2858082}
\showDOI{\tempurl}


\end{thebibliography}

\appendix
\section{Appendix A: Design Space Scoring Table}
\begin{figure}[H]
  \centering
  \includegraphics[width=1.0\linewidth]{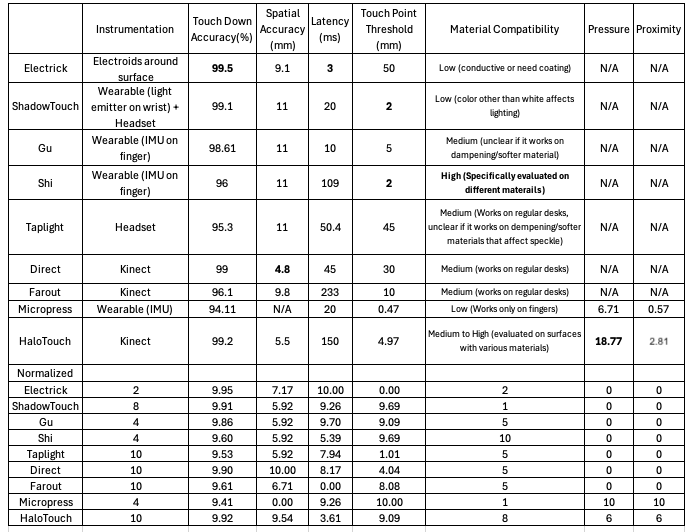}
  \caption{Comparison of Touch Input Systems Across Key Metrics. This table compares various touch input systems based on key performance metrics, including touch-down accuracy, spatial accuracy, latency, touch point threshold, material compatibility, pressure sensing, and proximity sensing. }
  \Description{A table comparing different touch input systems, including Electrick, ShadowTouch, Gu, Shi, Taplight, Direct, Farout, Micropress, and HaloTouch. The table presents metrics such as instrumentation type, touch-down accuracy (percent), spatial accuracy (millimeters), latency (milliseconds), touch point threshold (millimeters), material compatibility descriptions, and whether the system supports pressure and proximity sensing. HaloTouch shows a touch-down accuracy of 99.2\%, spatial accuracy of 5.5 mm, latency of 150 ms, and a touch point threshold of 4.97 mm. It also supports pressure sensing (18.77) and proximity sensing (2.81), unlike most other systems.}
  \label{fig:radar_table}
\end{figure} 

\section{Appendix B: Halo Effect Signal Strength}
\label{sec:halo_signal_strength}
\begin{figure}[H]
  \centering
  \includegraphics[width=1.0\linewidth]{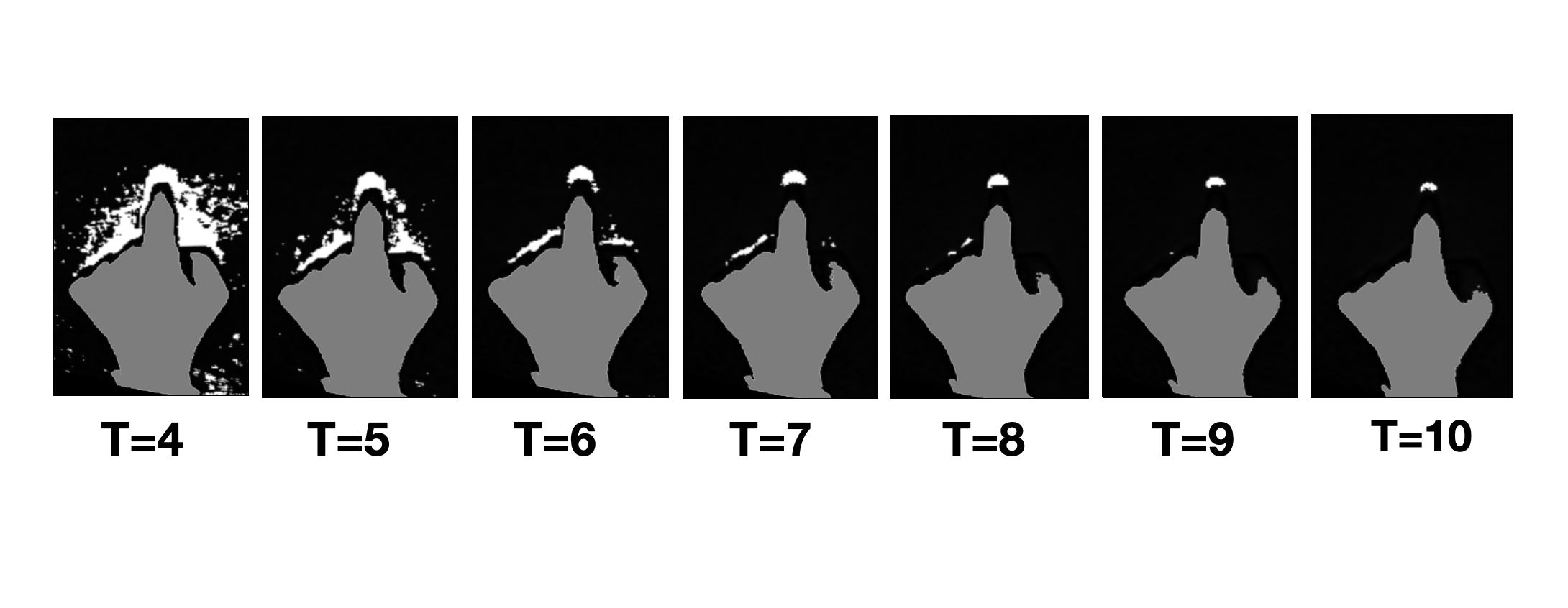}
  \caption{Effect of Threshold Variation on Halo Signal Strength. This figure illustrates the impact of different threshold values (T=4 to T=10) on the detection of the halo effect around a fingertip in depth camera imagery.}
  \Description{A sequence of seven grayscale images showing a hand with a pointing finger, labeled with threshold values from T=4 to T=10. Each image visualizes the halo effect detected by a depth camera, where bright regions indicate the signal strength around the fingertip. As the threshold increases, the noise surrounding the hand diminishes, making the halo effect more distinct and localized around the fingertip.}
  \label{fig:halo_signal_strength}
\end{figure} 

\begin{figure}[H]
  \centering
  \includegraphics[width=0.8\linewidth]{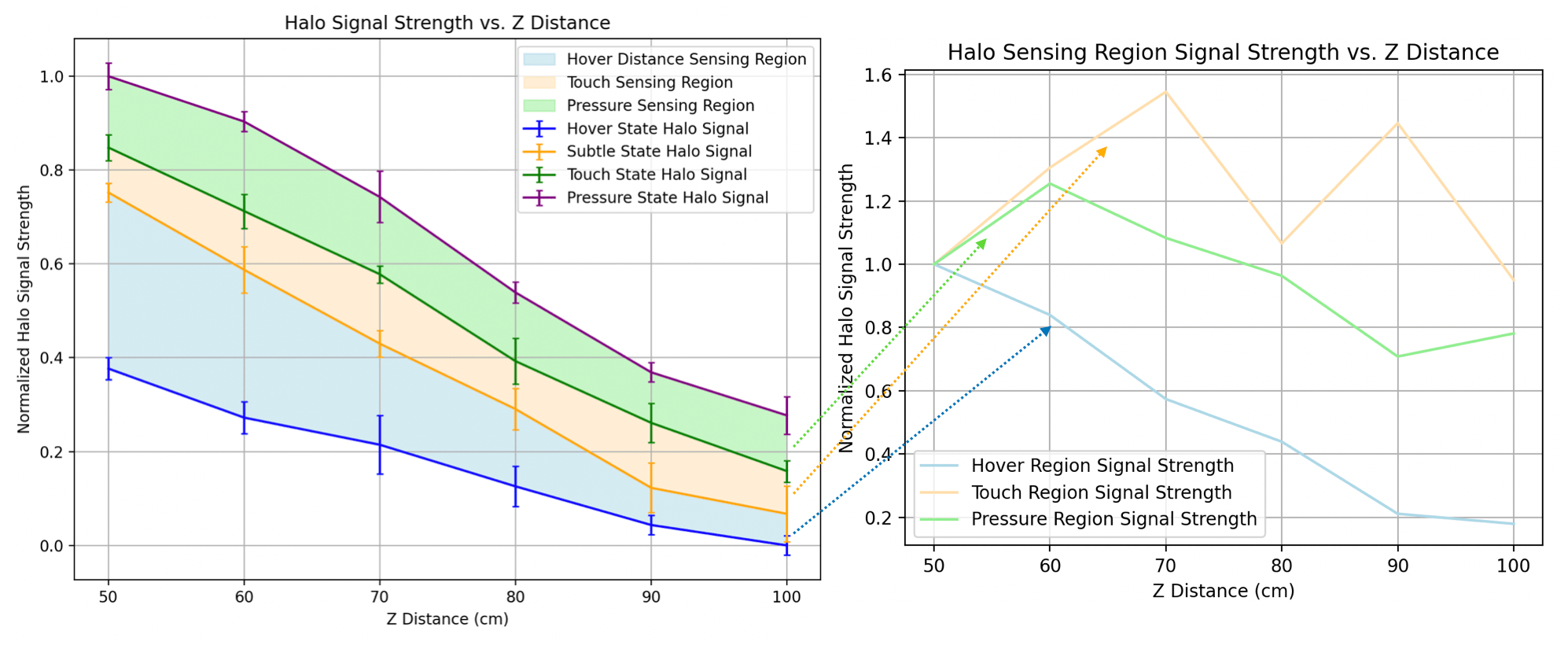}
  \caption{Halo Signal Strength Variation Across Z-Distance. These plots illustrate the relationship between Z-distance (distance from the sensor) and normalized halo signal strength for different interaction states (hover, touch, and pressure). The left plot shows signal strength degradation with increasing Z-distance, with shaded regions indicating different sensing zones. The right plot highlights how signal strength transitions across the halo sensing regions.}
  \Description{Two line graphs showing the relationship between Z-distance (in cm) and normalized halo signal strength. The left graph displays four signal curves for different interaction states (hover, subtle, touch, and pressure), with shaded regions representing hover, touch, and pressure sensing areas. The right graph visualizes halo sensing region signal strength trends, illustrating the changes in hover, touch, and pressure signals across distances. Error bars are included in the left graph to indicate variability. Both graphs highlight how halo signal strength diminishes with increasing Z-distance, affecting detection accuracy.}
  \label{fig:signal_strength_z}
\end{figure} 

\begin{figure}[H]
  \centering
  \includegraphics[width=0.8\linewidth]{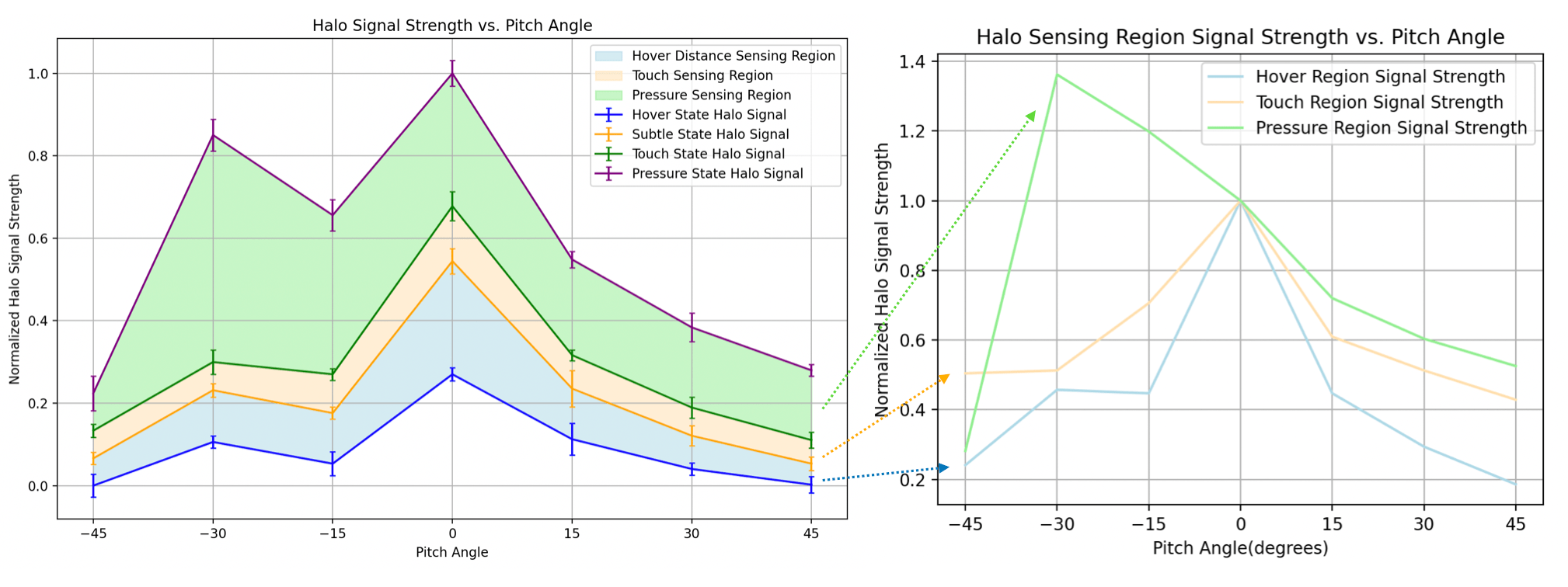}
  \caption{Halo Signal Strength Variation Across Pitch Angles. These plots illustrate the relationship between pitch angle and normalized halo signal strength for different interaction states (hover, touch, and pressure). The left plot shows how signal strength changes with pitch angle, where the shaded regions indicate hover, touch, and pressure sensing zones. The right plot focuses on the transition of signal strength across these sensing regions.}
  \Description{Two line graphs showing the relationship between pitch angle (in degrees) and normalized halo signal strength. The left graph presents four signal curves corresponding to different interaction states (hover, subtle, touch, and pressure), with shaded regions marking hover, touch, and pressure sensing areas. The right graph highlights the halo sensing region signal strength trends, depicting how hover, touch, and pressure signals change across varying pitch angles. Error bars in the left graph indicate variability. The data shows that halo signal strength is highest at 0° pitch angle and decreases symmetrically in both directions, affecting interaction accuracy.}
  \label{fig:signal_strength_pitch}
\end{figure} 

\section{Appendix C: Typing UER and CER}
\label{sec:typing_error}
\begin{figure}[H]
  \centering
  \includegraphics[width=0.85\linewidth]{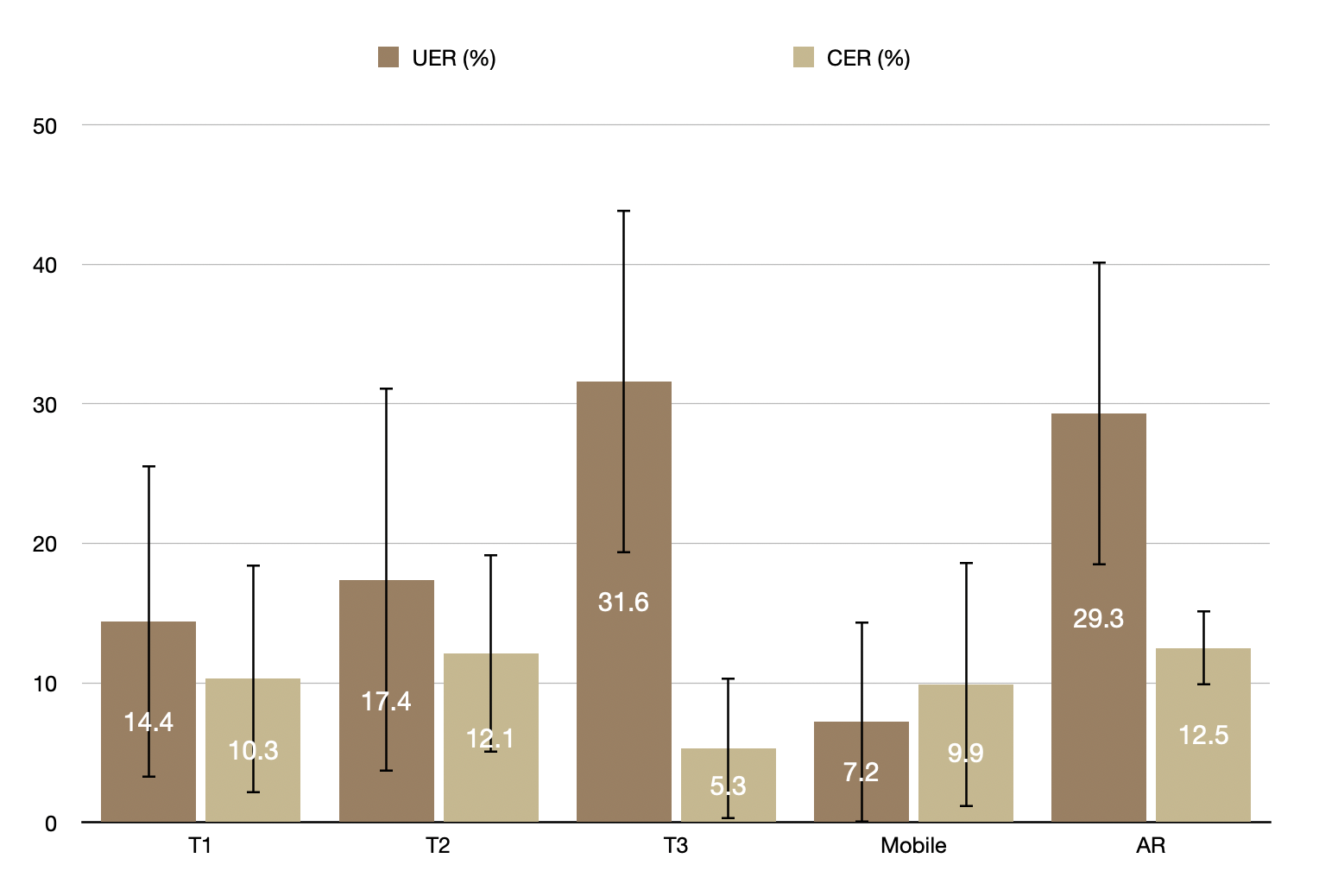}
  \caption{User Error Rate (UER) and Character Error Rate (CER) across different input methods. This bar chart compares UER (\%) and CER (\%) across five input conditions: T1, T2, T3, Mobile, and AR. Darker bars represent UER, while lighter bars represent CER, with values displayed inside each bar.}
  \Description{A bar chart comparing User Error Rate (UER\%) and Character Error Rate (CER\%) across five input conditions: T1, T2, T3, Mobile, and AR. The UER bars (darker) and CER bars (lighter) show corresponding percentages, with error bars indicating variability. The highest UER is observed in T3 (31.6\%) and AR (29.3\%), while the lowest error rates are in the Mobile condition (UER: 7.2\%, CER: 5.3\%). The results highlight significant differences in typing accuracy across these conditions.}
  \label{fig:typing_accuracy_error}
\end{figure} 
\end{document}